\PassOptionsToPackage{table}{xcolor}
\documentclass[manuscript,screen,anonymous=false,acmsmall]{acmart}

\usepackage{booktabs}  
\usepackage[table,xcdraw]{xcolor}
\usepackage{listings}
\usepackage{graphicx}
\usepackage{tabularx}
\usepackage{adjustbox}
\usepackage{enumitem}
\usepackage{pifont}
\usepackage{tikz} 
\usepackage{lipsum} 
\usepackage{multirow}

\newcommand{\customitem}[1]{\item[\ding{#1}]}

\newenvironment{customitemize}
{ \begin{itemize}[left=0pt, itemsep=0.5em] }
{ \end{itemize} }

\lstdefinestyle{mystyle}{
    basicstyle=\ttfamily\small,
    backgroundcolor=\color{lightgray},
    frame=single,
    breakatwhitespace=true 
}

\AtBeginDocument{%
  \providecommand\BibTeX{{%
    \normalfont B\kern-0.5em{\scshape i\kern-0.25em b}\kern-0.8em\TeX}}}

\setcopyright{acmlicensed}
\copyrightyear{2025}
\acmYear{2025}
\acmDOI{XXXXXXX.XXXXXXX}

\acmConference[XXX]{Make sure to enter the correct
  conference title from your rights confirmation email}{October 18--22, 2025}{XXX, XXX}
\acmISBN{978-1-4503-XXXX-X/18/06}

\begin{document}

\title[Collecting Social Data with Large Language Models]{Collecting Qualitative Data at Scale with Large Language Models: A Case Study}

\author{Alejandro Cuevas}
\email{acuevasv@cs.cmu.edu}
\affiliation{%
  \institution{Carnegie Mellon University}
  \streetaddress{5000 Forbes Ave.}
  \city{Pittsburgh}
  \state{Pennsylvania}
  \country{USA}
  \postcode{15201}
}
\authornote{The author completed this work as part of an internship at Microsoft Research.}
\orcid{0000-0001-6507-1334}

\author{Jennifer V. Scurrell}
\affiliation{%
  \institution{ETH Zurich}
  \city{Zurich}
  \country{Switzerland}}
\email{jennifer.scurrell@sipo.gess.ethz.ch}

\author{Eva M. Brown}
\email{evamxb@uw.edu}
\affiliation{%
  \institution{University of Washington}
  \city{Seattle}
  \state{Washington}
  \country{USA}
}

\author{Jason Entenmann}
\affiliation{%
  \institution{Microsoft Research}
  \city{Redmond}
  \state{Washington}
  \country{USA}
}

\author{Madeleine I. G. Daepp}
\affiliation{%
 \institution{Microsoft Research}
 \city{Redmond}
 \state{Washington}
 \country{USA}}

\renewcommand{\shortauthors}{Cuevas et al.}

\begin{abstract}
  Chatbots have shown promise as tools to scale qualitative data collection.
  Recent advances in Large Language Models (LLMs) could accelerate this process
  by allowing researchers to easily deploy sophisticated interviewing chatbots.
  We test this assumption by conducting a large-scale user study (n=399)
  evaluating 3 different chatbots, two of which are LLM-based and a baseline
  which employs hard-coded questions. We evaluate the results with respect to
  participant engagement and experience, established metrics of chatbot quality
  grounded in theories of effective communication, and a novel scale evaluating
  ``richness'' or the extent to which responses capture the complexity and
  specificity of the social context under study. We find that, while the
  chatbots were able to elicit high-quality responses based on established
  evaluation metrics, the responses rarely capture participants' specific
  motives or personalized examples, and thus perform poorly with respect to
  richness. We further find low inter-rater reliability between LLMs and humans
  in the assessment of both quality and richness metrics. Our study offers a
  cautionary tale for scaling and evaluating qualitative research with LLMs.

\end{abstract}

\ccsdesc[500]{Human-centered computing~Empirical studies in collaborative and social computing}

\keywords{Chatbots, Qualitative Research, Language Models, GPT}

\received{16 January 2024}
\received[revised]{16 July 2024}
\received[accepted]{20 September 2024}

\maketitle

\section{Introduction}

Improvements in the collection and analysis of qualitative data can help bridge
the gap between quantitative and qualitative methods. Social scientists who seek
to understand people's experiences and perspectives frequently face trade-offs
between the scale and speed of quantitative approaches to data collection and
the richness of qualitative methods~\cite{small2022qualitative,
leedy2015practical, lazar2017research}. Interviewing, the hallmark method of
qualitative research, is a time- and labor-intensive process. Thus, many
interview studies remain small in scale~\cite{caine2016local} and those studies
done at large scales can take years~\cite{hunt2011embarking}---meaning that HCI
and CSCW researchers working in fast-paced and rapidly changing technological
environments can struggle to provide timely insights.

Advances in computational methods, specifically Artificial Intelligence (AI),
have introduced many methodological efficiencies to assist qualitative research.
These improvements can be divided by whether they affect the evaluation or the
collection of qualitative data. In terms of evaluation, many improvements have
focused on reducing the manual burden involved in qualitative analysis, such as
by facilitating text annotation~\cite{13-dam-kaufmann-2008-computer}, extracting
topics from open
responses~\cite{13-dam-kaufmann-2008-computer,1-bakharia-etal-2016-interactive},
helping in the generation of codebooks and coding
data~\cite{2018automated_coding,11-chen-etal-2016-challenges}. In terms of
collection, AI-driven chatbots have enabled the collection of data at
scale~\cite{xiao2020, xiao2019, jiang2023communitybots,xiao2023}, mostly in
conversational studies, but also in ethnographic studies~\cite{ethnobot}. The
advent of highly capable Large Language Models (LLMs) has introduced remarkable
improvements across the aforementioned
tasks~\cite{gilardi2023chatgpt,xiang_llm_coding}, while also enabling new use
cases, like creating synthetic user research data~\cite{synthetic_llm}. 

Most notably, the advent of LLMs has made deploying highly sophisticated
chatbots remarkably easy. Before LLMs, chatbot adoption remained limited due to
the brittleness of hybrid rules-based and machine learning-based
systems~\cite{grudin2019}, with unexpected limitations and unpredictable failure
modes~\cite{ashktorab2019resilient}. As a result, chatbot interactions
frequently failed to meet user expectations~\cite{luger2016}. However, LLMs
promise to enable new ease of deployment and improvements in cognitive
capabilities that could foster vast improvements in the flexibility and
extensibility of chatbot-based tooling.

An important open question is whether these models can scale the collection of
\textit{high-quality and rich qualitative data}. The current literature suggests
that existing conversational agents are suitable for qualitative studies and can
elicit high-quality
data~\cite{xiao2019,xiao2020,jiang2023communitybots,chopra2023conducting}. The
evaluation of quality in chatbots has ranged from quantitative measures (e.g.,
word overlaps between text corpora)~\cite{quant_measures_chatbots} to
mixed-methods assessments of effective communication~\cite{xiao2019, xiao2020,
jiang2023communitybots}. In particular, a recent body of work has employed
quality evaluations via human annotation~\cite{xiao2019, xiao2020,
jiang2023communitybots}. These evaluations and metrics are based on Gricean
Maxims, a ``set of communication principles that a speaker and listener should
adhere to ensure effective communication''~\cite{xiao2020,grice1975logic}. But
longstanding debates amongst scholars of qualitative research suggest that
interactions that perform well on conventional and quantitative measures of
quality can still fall short of constituting ``good'' qualitative
research~\cite{king2021designing, mahoney2006tale, mahoney2010after,
small2022qualitative}. While effective communication is an essential component
of data collection, the metrics above may be insufficient at capturing the
richness of the data: the extent to which research captures the specificity and
complexity of the social context being studied~\cite{small2022qualitative}.

To address this gap, we operationalize new metrics to evaluate the richness of
chatbot conversations. We anchor our approach on the indicators proposed by
Small and Calarco ~\cite{small2022qualitative}---a framework for assessing
richness emerging as a foundational text in qualitative social
science~\cite{Conzon2023, Epic2024}. We leverage Small and Calarco's work
throughout our chatbot development and evaluation process. First, we ground the
development of two LLM-based modules around key metrics of cognitive empathy
(understanding participants' motivations and value systems) and palpability
(surfacing specific, personal evidence), resulting in a chatbot that dynamically
generates follow-ups to participants' responses. We then use these chatbots to
conduct a large-scale (n=399) user study on AI alignment, which enables us to
compare both conventional measures of quality and our novel richness metrics
across LLM-augmented versus a baseline of naive interactions. Finally, we
compare inter-rater reliability for an LLM versus humans in assessing
interactions' quality and richness. Through this study, we make four
contributions to the CSCW literature.

\begin{itemize}
    \item We showcase the ease of deployment of an LLM-augmented chatbot.
    Through the development of two modules, we create a chatbot that can
    dynamically generate follow-up questions and provide users with real-time
    conversation summaries. We provide an open-source codebase and prompts to
    enable other CSCW researchers to use and expand upon our research.
    \item We develop a novel richness scale grounded in social scientific
    principles and demonstrate that existing quality metrics do not indicate
    richness. 
    \item We document common interviewing mistakes made by LLMs, including the
    emergence of bias both in responses and in participant treatment of the bot,
    and demonstrate how these mistakes inhibit using chatbots for automating
    qualitative research.
    \item We show that even state-of-the-science LLMs struggle to consistently
    evaluate data quality and response richness indicators, which may explain
    their current limitations in generating follow-ups that elicit rich
    responses.
\end{itemize}


\section{Related Work}
\label{sec:related-work}

Our study is informed by Human-Computer Interaction (HCI) research
methodologies, prior work on the design of information elicitation chatbots,
theoretical understanding of human-chatbot interactions, and empirical and
theoretical work on response evaluation in qualitative research.

\subsection{Evaluation Metrics for Chatbots}

Chatbots have typically been evaluated on three dimensions: quantitative,
user-oriented, and qualitative text-related measures. Quantitative measures
encompass metrics such as the number of positive negative
words~\cite{74kawasaki2020assessing}, the number of
words~\cite{54foster2009comparing}, the number of personal pronouns
used~\cite{166thomas2020expressions}, and embedding-based metrics such as
corpora overlap~\cite{quant_measures_chatbots}. These metrics are often used in
scale analyses, as they are computed and thus often easily
scalable~\cite{quant_measures_chatbots}. Similarly, user-centric metrics also
mainly encompass easily scalable metrics, such as completion time, task
fulfillment, dropout rate~\cite{45demberg2011strategy,131pecune2018field}, and
self-disclosed satisfaction
metrics~\cite{45demberg2011strategy,97liu2020cold,xiao2020,
paradise,zarouali2023comparing}. 

Qualitative metrics, on the other hand, have only recently received more
attention. Evaluations of ``quality'' have taken many forms, depending on the
context of the chatbot deployment~\cite{chatbot_paper_review}. The metrics
proposed by Xiao et al., which evaluate communicative quality, are the most
relevant. These metrics are based on the Gricean maxims: four principles of
effective communication including quality, quantity, relevance, and
communication manner~\cite{grice1975logic}. Xiao et al. operationalize these
ideas through measures of response, informativeness, specificity, relevance, and
clarity~\cite{xiao2020,xiao2019,xiao2023}, finding that the chatbot offered
higher quality responses in terms of relevance, specificity, and clarity
compared to the responses collected by a web-based
survey~\cite{xiao2020,xiao2023}. Jiang et al. use the same metrics to assess a
multi-agent chatbot architecture~\cite{jiang2023communitybots}. 

Despite the high performance reported by recent chatbots in many of the quality
metrics stated in the chatbot literature~\cite{chatbot_paper_review}, we have
not seen widespread adoption of these tools for data collection--suggesting a
gap between the measured and actual usefulness of the data gained. We seek to
explain this gap by comparing measures currently used in the literature with a
novel set of measures used to evaluate qualitative research.

\subsection{Evaluating Qualitative Research Through \textit{Richness}}

Although there is no established set of metrics used to evaluate qualitative
data~\cite{mahoney2006tale,small2022qualitative}, various researchers have
proposed quality indicators such as rich rigor~\cite{tracy2010qualitative},
thickness~\cite{geertz1973interpretation}, and
palpability~\cite{saldana2021coding,small2022qualitative}. Patton writes that
producing valuable qualitative insight requires surfacing ``direct quotations
from people about their experiences, opinions, feelings, and
knowledge"~\cite{patton2014qualitative}. Geertz describes a detailed description
that makes evident the meanings and significance behind the participants'
responses as ``thick''~\cite{geertz1973interpretation}. That is, interviewees'
responses should not merely be on topic and comprehensive but provide tangible
insights specific to individual participants that enable the researcher and,
ultimately, the reader to better understand the world from the participant's
perspective. The evaluation of the \textit{quality} of qualitative research,
then, requires an assessment of the extent to which responses capture insights
from people that are \textit{rich} enough to allow a detailed and nuanced
understanding of the personal experiences of the participants in the condition
under study~\cite{patton2014qualitative, merriam2015qualitative,
firestone1987meaning, geertz1973interpretation, denzin2011sage}.

To provide a comprehensive set of indicators to evaluate qualitative research,
Small and Calarco~\cite{small2022qualitative} define five key constructs that
this broader literature on qualitative methods identifies as critical to
``quality'' in qualitative research. We leverage this crucial theoretical
contribution to introduce a novel set of quality measures grounded in theory and
research based on principles of effective qualitative research. We call these
metrics \textit{measures of ``richness''} to reflect the extent to which they
seek to evaluate the depth of understanding that responses make possible for the
researcher.

Specifically, we operationalize four quality metrics identified in Small and
Calarco's textbook as indicative of effective qualitative research: (1)
cognitive empathy, or the researcher's ability to understand participant
motives; (2) palpability, the inclusion of concrete and personal evidence
(rather than abstractions of generalizations); (3) follow-up, the extent to
which the researcher collects data related to new questions as they arise
through the data collection process; and (4) self-awareness, or the
interviewers' understanding of the impact of their identity on those
interviewed, and thus on the collected data\footnote{Small and Calarco also
describe the role of heterogeneity in obtaining sound qualitative results but
note that ``[heterogeneity] describes not a feature of the research design but
the character of the written report''\cite{small2022qualitative}, and as such
heterogeneity is omitted from our coding guide.}. We show how responses that
perform well on measures of communicative effectiveness remain sparse in terms
of richness. Thus, these metrics help to explain how open-ended chatbot answers
that are relevant, specific, clear, and informative can nevertheless fall short
of the quality of data an effective human interviewer would collect.

\subsection{Eliciting High-quality Through Design and Cognition}

Design choices, such as minor changes to user interfaces, can also have
measurable effects on evaluations. For example, Candello et al. found quality
gains simply by varying the choice of font~\cite{Candello2017}. Kim et al. found
that moving from a web-based survey to a chatbot-based conversation can improve
user engagement, response quality, ease of use, and user enjoyment of the survey
experience~\cite{kim2019}. However, the choice of chatbot along with its
``style, features, and affordances'' need to be considered to surpass the
experience of using a web-based survey~\cite{zarouali2023comparing}.

Cognitively, the ``personality'' of chatbots, as well as the use of simple
humanization techniques such as conducting self-introductions, using
respondents' names, or echoing the responses of the participants, seem to
significantly affect user engagement, trust, and perceived
experience~\cite{zhou2019,li2017confiding}. Xiao et al.~\cite{xiao2019}
developed a chatbot for text-based interview tasks that can ``actively listen''
or produce on-topic replies to participant responses. They later expanded on
this work by implementing an open-ended conversational chatbot-based survey that
``probes'' by asking follow-up questions based on the content of the
participant's responses~\cite{xiao2020}. They observe that these augmentations
measurably improve engagement length and responses' relevance and specificity
compared to an open-ended survey~\cite{xiao2020} or a baseline chatbot without
active listening skills~\cite{xiao2019}.

Beyond personality and humanization, knowledge and the ability to handle
conversations seamlessly are important cognitive traits for chatbots. In the
study conducted by Xiao et al., participants expressed that specific responses
were ``vague and shallow,'' which they believe could be addressed if the chatbot
can extract ``deeper user intent~\cite{xiao2019}.'' In a recent study, Jiang et
al. leverage a multi-agent architecture to reduce interruption and create more
engaging conversations~\cite{jiang2023communitybots}. These domain-specific and
multi-agent chatbots are used for eliciting public input, and they find them to
be more effective than a single-agent
architecture~\cite{jiang2023communitybots}. These studies support the need for
flexibility and topic generalizability in chatbot design. We contribute to this
literature by exploring the effect of enhanced cognitive capabilities
(facilitated by an LLM) on participant experience, data quality, and response
richness.

\section{Chatbot Development}

In developing our chatbot, we created two LLM-based modules grounded in the
methods of effective interviewers~\cite{small2022qualitative}. Our ``north
star'' was to surface rich qualitative data. In human interviews, attaining
richness requires thoughtful follow-up questioning or ``probes'' on the part of
the interviewer~\cite{small2022qualitative}. We thus designed an LLM-based
``Dynamic Prober'' module that allows the chatbot to generate follow-up
questions based on participants' responses. Second, recognizing the value of
self-awareness~\cite{small2022qualitative}---the importance of addressing biases
in the LLM's assessment of the conversation---we created a ``Member Checker''
module that enabled the participant to confirm or contest a chatbot-generated
summary of the interaction, as in interview and participatory studies in which
researchers seek to confirm results by debriefing
participants~\cite{morse2002verification}. An overview of the architecture is
included in the Appendix (Figure~\ref{fig:systemdiagram}). 

We developed each agent through the prompting of OpenAI's GPT-based LLMs.
Prompting is emerging as the leading approach for flexible development of
conversational agents~\cite{liu2023pre}, but writing effective prompts is a
nontrivial task~\cite{zamfirescu2023johnny}. There do not yet exist established
workflows for systematic and robust prompt design, and novice prompters tend to
take ad-hoc approaches that result in brittle, limited
solutions~\cite{zamfirescu2023johnny}. In this section, we describe our
three-step process to systematize the development of prompts for each module,
including: identifying relevant design principles from the natural language
processing literature; systematically refining prompts grounded in these
principles via iteration on synthetic datasets; and improving robustness through
external red-teaming of the final modules. We also delineate the key theoretical
and practical considerations underlying our approach to the user interface
design. Our chatbot and prompts are available at: \url{https://github.com/microsoft/dynamic-prober}

\begin{figure}
    \centering
    \includegraphics[width=\linewidth]{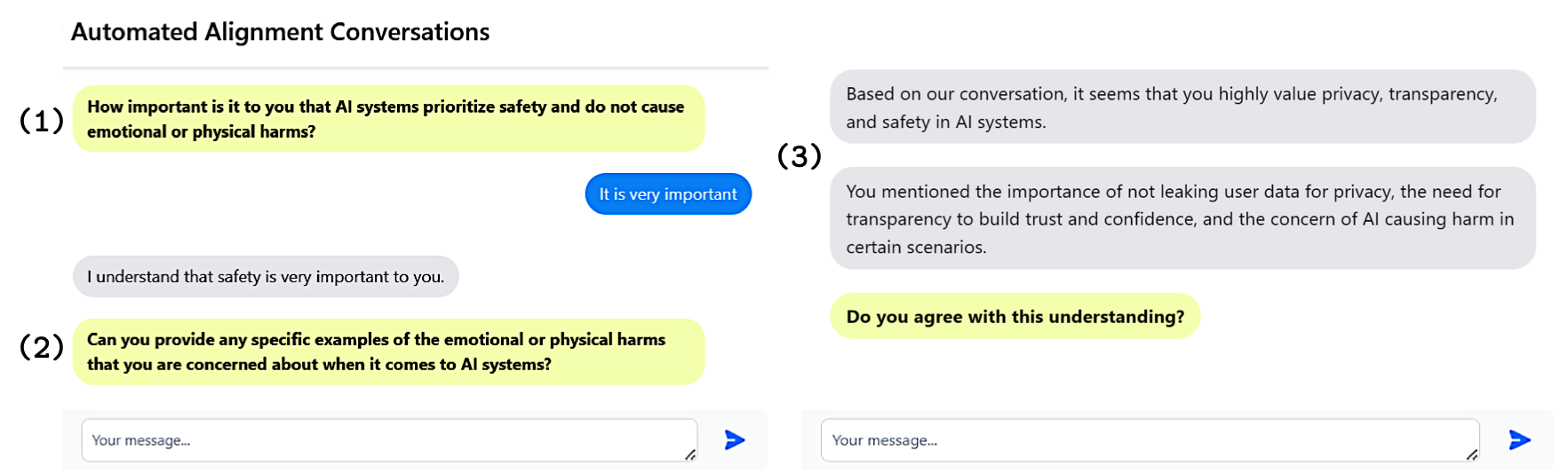}
    \caption{Chatbot User Interface. Generated text is indicated in grey with
    questions made more salient to the user through yellow highlighting. User
    text is indicated in blue. The image presents an examples from a dialogue in
    which (1) the chatbot first showed a question from an established, validated
    measurement scale~\cite{jakesch2022different}, modified to use less formal
    language~\cite{kim2019} and then (2) accessed the ``Dynamic Prober'' module
    to generate a follow-up question based on the user's response. After the
    user responded to three questions and their associated follow-up probes, (3)
    shows the ``Member Checker'' module in which the chatbot generated a summary
    of the conversation and asked the user to confirm the summary content.}
    \label{fig:chatbot_ui}
    \Description{Two screen captures from our a conversation with our chatbot
    application. The layout is that of message chat bubbles. On the left side, a
    question is provided and marked with a (1) to indicate that it's pulled from
    hard-coded questions. Questions are highlighted in yellow. Participant
    responses are in blue. The bot provides a follow-up comment, colored gray,
    and a question highlighted in yellow. These are marked as (2). Lastly, on
    the right side, we see the chatbot's summary of the conversation marked with
    a (3), followed with a agree/disagree question for the participant.}
\end{figure}

\subsection{Design Principles for Effective Prompting}
\label{subsec:llm_modules}

We used a prompt engineering approach in which we created multiple agents each
with their own prompt (a set of instructions pre-pended to user interactions
with the LLM) and adaptable context (an updated section of the conversation to
which they had access and could respond). We chose not to use more complex
design approaches such as Retrieval Augmented Generation (RAG) given the limited
knowledge retrieval and relatively simple dialogue structure required for an
effective interview, as well as concerns over latency in our setting. Although
there is not yet a standardized methodology for effective prompt
development~\cite{zamfirescu2023johnny}, researchers have uncovered a number of
``prompt engineering'' methods that can improve and stabilize LLM performance.
At the time of development, the following strategies were the
state-of-the-science approaches to effective prompting.

\begin{customitemize}
    \customitem{182}  \textit{Chain-of-thought}: follow intermediate reasoning
    steps before producing  final output~\cite{wei2022chain, kojima2022large}.
    \customitem{183}  \textit{Few-shot examples}: including a small number of
    examples of the desired output~\cite{brown2020language}.
    \customitem{184}  \textit{Structured output}: produce a response in a
    structured data format such as a JSON list~\cite{white2023prompt}.
    model training data contains a large fraction of structured data, results
    can be more stable and the failure to use a format provides an automated
    debugging check
    \customitem{185}  \textit{Generated-knowledge prompting}: generate relevant
    additional information before completing a prompt, ensuring that the
    completion is conditioned on the additional
    information~\cite{liu2021generated}.
    after each of the others also?
    \customitem{186} \textit{Role assignment}: initializing the system with a
    message that primes the model with a role such as a job or personality,
    which can provide contextual information relevant to model
    performance~\cite{white2023prompt}.
\end{customitemize}

We grounded our development of the modules in these approaches, iteratively
implementing and testing these approaches on dummy synthetic data, evaluating
the improvement with generated synthetic data, and then conducting a final
evaluation via ``red-teaming'' by human testers.

\subsection{Systematically Refining Prompts with Synthetic Data}

Recent work has found that LLMs can aid in piloting and testing HCI instruments
by generating synthetic research data~\cite{synthetic_llm}. To develop modules
that could robustly produce relevant probes and effectively summarize
conversations, we created two synthetic data sets with which to (1) iteratively
test and develop our prompts (2) evaluate the robustness and stability of the
final prompts. Lastly, we conducted an initial pilot with external colleagues.

\subsubsection{Iterative Prompt Development}

In our study, we employ seven questions on AI alignment drawn
from~\cite{jakesch2022different}. In each chatbot interaction, we planned to
randomly pick three questions to discuss with participants (this procedure is
described in Section~\ref{sec:user_study}). Thus, to begin the development of
our prompt, we created a synthetic dataset consisting of individuals with the
following attributes: an importance score and verbal for each value in the
survey, a job, and an explanation about their role and motive (e.g.,``Not
important. As a software engineer, I don’t care about fairness. I think
performance is the most important priority for AI alignment.''). Using these
explanations, we attempted to generate suitable follow-ups (e.g., ``I see that
fairness is not as important as performance to you. Can you think of an example
in software engineering where fairness is dependent on performance?''). While
the initial follow-ups contained simple examples, they provided a starting point
to streamline and systematize the development of the first prompts.

We then generated a second more comprehensive synthetic data set for more robust
evaluation (see Appendix~\ref{apx:synthetic_descriptions}, for descriptions of
the attributes we used). In our study, our LLM would not necessarily have access
to these data, unless participants self-disclosed in the conversation. However,
including demographic and political traits in the synthetic data set was useful
to generate more nuanced responses and richer dialogs, which allowed us to
explore more edge cases and potential failure modes in the conversation. This
time, rather than just using the explanations as inputs, we use an LLM to
simulate a conversational interaction similar to H\"am\"al\"ainen et
al.~\cite{synthetic_llm}. We then prompted GPT-4 to review the participant's
profile and position on one of the survey questions, to develop a motive for
their perspective, and to simulate responses accordingly. For ease of
evaluation, each persona prompt completed one answer to an interview question,
one interaction with the ``Dynamic Prober'' module, and one interaction with the
``Member Checker'' module.

\subsubsection{Prompt Evaluation}

We assessed the resulting conversations qualitatively. First, we reviewed the
simulated conversations to assess the robustness and quality of the probes
produced by the ``Dynamic Prober'' module and the summaries by the ``Member
Checker'' Module (assessing follow-up and self-awareness, respectively, per
Table~\ref{tab:indicator-defs}). Second, we applied a ``coder'' module that
sought to surface (1) the importance score, as a quantitative measure of
accuracy; (2) the motive (as a qualitative measure of palpability and cognitive
empathy~(Table~\ref{tab:indicator-defs}); and (3) and an explanation for the
bot's results, to aid in debugging. Qualitative review of the conversations
further found that, unlike initial prompts, our final module prompts produced
stable responses in every interaction, including the bad-faith conversation.
Following common practices for red teaming AI systems~\cite{ganguli2022red,
xu2021bot}, we then asked colleagues external to the research team to test the
final experience.

\subsubsection{Effectiveness of Prompt Strategies}

We observed improvements in both robustness and quality through the use of
prompt design principles. First, the use of \ding{182} \textit{chain-of-thought}
prompting enabled the ``Dynamic Prober'' module to check for bad-faith
interactions before producing a response, and improved both modules'
faithfulness to the prompt. The inclusion of \ding{183} \textit{few-shot
examples} similarly improved the ``Dynamic Prober'' module's ability to
recognize and deflect bad-faith responses and fostered a consistent tone and
response style across modules. Further, the \ding{184} \textit{structured
output} format reduced failures across both modules by enabling automated type
checking. From an engineering perspective, it also had the benefit of ensuring
consistent formatting and thus facilitating additional computation on the
results. We also achieved improvements by procedurally generating the output
(\textit{\ding{185} generated-knowledge}). In early prompts, we simply asked the
model to generate follow-up questions based on the context or to generate
multiple possible follow-up questions and then to select the ``best''. We found
that the model frequently created off-topic or low-quality, albeit valid,
questions. However, when we prompted the model to produce a JSON list comprising
the user's opinion, the user's given reasoning for that opinion (if any), and to
identify unanswered aspects of an opinion before generating its follow-up
question, we observed that the resulting probes were consistently conditioned on
the prior information and thus more relevant to the conversation. Only
\ding{186} \textit{role assignment} failed to noticeably impact the quality of
questions. We did not observe substantial differences when priming the model to
be, for example, a ``professional interviewer'', or ``an expert qualitative
researcher'', and thus we did not assign roles in our final prompts.

Overall, the evidence produced with the synthetic data set as well as the
experiences of human testers bolstered our confidence in the robust performance
of our prompts both in good- and simple bad-faith interactions. We note,
however, that participants who sought to derail the conversation were able to do
so with enough attempts and ingenuity. Developing prompts that can effectively
thwart adversarial attempts remains a difficult problem, as observed by the
ongoing efforts of companies and researchers to develop robust
models~\cite{llamaguard2023,openaiguardrails2024}, as well as ongoing efforts to
defeat their defenses~\cite{crescendoattack2024,jailbreakattacks2024}. Because
our research population comprised people in a professional setting, we assumed
that most participants would act in good faith. We caution researchers who seek
to deploy LLM-based chatbots that this assumption may not hold true in other
settings.

\subsection{Chatbot Interface Design}\label{chatbot:design}

Our chatbot had a simple conversational interface (Figure~\ref{fig:chatbot_ui}).
However, we made four key design choices informed by prior literature. First,
given evidence that conversational style meaningfully and significantly affects
user engagement and experience~\cite{kim2019}, two members of the research team
collaborated to revise the primary questions to use a more casual language.
Second, we implemented a rule-based turn-taking approach in which users were
only able to submit a single response to a chatbot-based question to ensure that
the conversation was chatbot-led and not human-initiated. Third, we included an
introductory message and question, consistent with research on the benefits of
humanization in chatbot interactions~\cite{humanizingchatbot}. In the ``Member
Checker'' condition, which was meant to most approximate an interview
experience, we also included a warm-up question based on evidence on the
importance of warm-up questions in interview research~\cite{warmupquestions}. We
additionally experimented with adding personalization in the follow-ups.
Personalization was done by asking the participant to share some information
about their role at the beginning of the interaction and then feeding that
information to the ``Dynamic Prober'' module. While this approach led to more
contextually-relevant questions, we had to disable it due to issues with staying
on topic. Fourth, leveraging feedback of early testers, we used color and text
to emphasize questions versus contextual statements. We also separated each
sentence into a separate speech bubble, rather than a large paragraph, and we
only allowed the chatbot to pose a single question in a given turn. These tweaks
aimed at improving the accessibility and legibility of the dialogue. Further
deployment details can be found in Appendix~\ref{apx:deployment}.

\section{User Study}
\label{sec:user_study}

We evaluated our chatbot through a large-scale user study with employees of a
major technology company. In our evaluation, we sought to disentangle the effect
of employing a sophisticated conversational agent versus a conversationally
interactive design without cognitive capabilities on response elicitation and
perception. Second, we sought to evaluate the quality and richness of the
responses in the study of AI alignment.

\begin{figure}
    \centering
    \includegraphics[width=\textwidth]{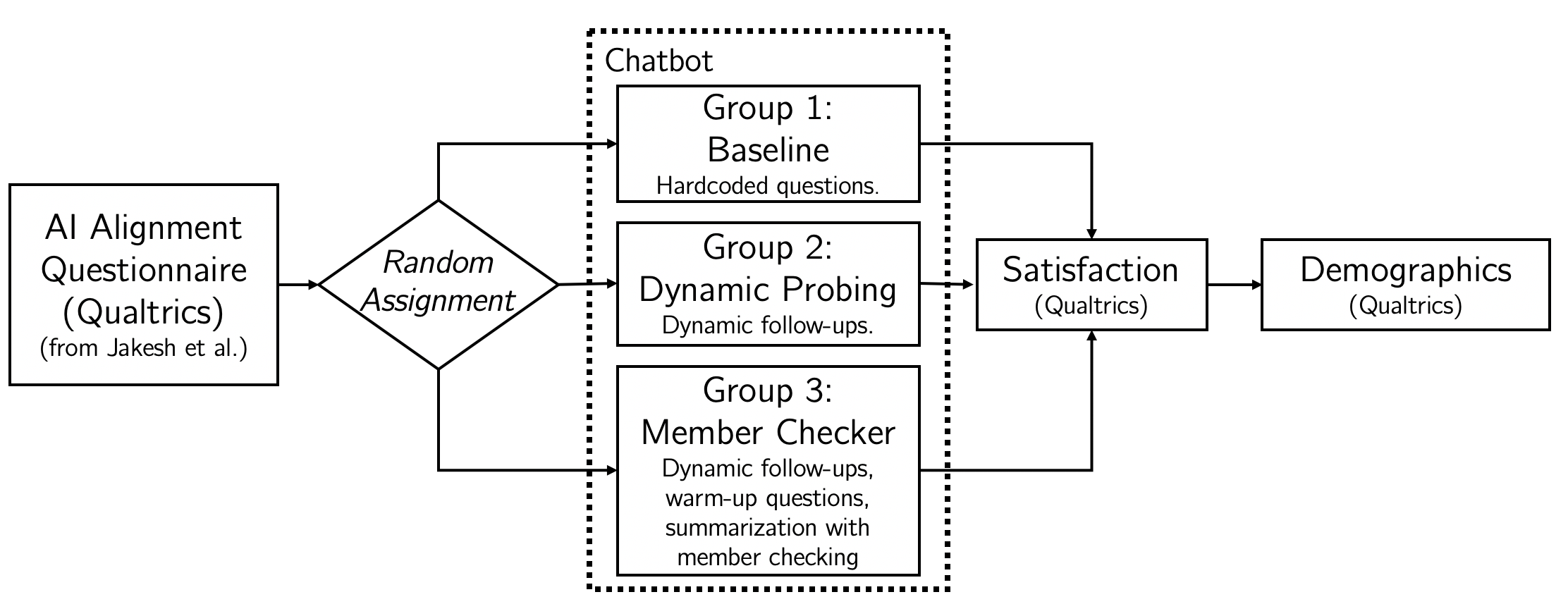}
    \caption{Experimental Design. Participants began the study by answering questions on AI alignment. They were then evenly randomized across three groups and taken to the Web-based chatbot interface. Group (1) was given hard-coded questions. Group (2) received the ``Dynamic Prober'' module. Group (3) interacted with the ``Dynamic Prober'' and ``Member Checker'' modules. All participants were then asked to complete questions on their experience and demographics.}
    \label{fig:studydesign}
    \Description{A flowchart showcasing our study design is presented. The
    flowchart begins with ``AI Alignment Questionnaire (Qualtrics) (from Jakesh
    et al.).'' An arrow then connects to a diamond shaped box labeled ``Random
    Assignment'', which in turns splits into three arrows, one for each group
    presented. Each group represents a possible chatbot treatment. Group 1 is
    the Baseline and has hard-coded questions. Group 2 engages with the
    ``Dynamic Prober'' and has dynamic follow-ups. Group 3 engages with the
    ``Member Checker'' which has dynamic follow-ups and summarization with
    member checking. Each of these groups then connects to a ``Satisfaction
    (Qualtrics)'' box. And lastly, the satisfaction box connects to
    ``Demographics (Qualtrics).''}
\end{figure}

\subsection{Study Topic} 

An aim of our work is to contribute to methodological innovations that support
CSCW and HCI research on rapidly changing technologies and in time-sensitive
contexts. In evaluating our prototype, we thus chose to conduct a study focusing
on a developing research area: AI Alignment, the identification of priority
areas and methods to ensure that AI systems act in ways consistent with human
values and goals~\cite{russell2010artificial}. 

\subsection{Study Design}\label{sec:userstudy}

We evaluated our prototype through a user study with US-based employees of a
large technology company. We conducted the user study in two waves: a small
initial pilot to test for errors followed by the main, larger deployment.
Figure~\ref{fig:studydesign} displays an overview of the study design. 

After giving their consent to participate, participants were shown definitions
of various AI values and asked to provide responses to a seven-question survey
on alignment hosted in a web-based form. These questions were drawn from a
previous study of AI alignment~\cite{jakesch2022different}, and their goal was
to prime participants with the definitions of the values that would be
discussed. Then each participant was given a unique link to the chatbot, where
they were randomly assigned to one of three conditions: (1) Baseline; (2)
``Dynamic Prober''; and (3) ``Member Checker''. 

The dialogue proceeded as follows. The chatbot first described the topic and
asked the user to confirm their readiness for the interaction. In the ``Member
Checker'' condition, which was meant to most approximate an interview, the
chatbot then posed an initial predetermined warm-up question (see
Section~\ref{chatbot:design}). After receiving the participant's response either
regarding readiness (Baseline and ``Dynamic Prober'') or on background (``Member
Checker''), the chatbot asked a randomly selected question from its
survey-derived list of primary questions and then ---after receiving the
participant's response---asked two questions. In the baseline condition, the
questions were always ``can you elaborate?'' and ``can you provide an example'',
regardless of the participant's input. In the conditions ``Dynamic Prober'' and
``Member Checker'', the questions were generated by the ``Dynamic Prober''. This
process was then repeated two times for a total interaction comprising three
question segments randomly selected from the original seven-question measurement
scale. At the end of the interaction, participants in the Baseline and ``Dynamic
Prober'' condition received a message thanking them for their time and directing
them back to the web-based survey platform with a user-specific completion code,
while in the ``Member Checker'' condition, the chatbot additionally provided
participants with a summary of the conversation and asked the participants to
confirm or contest the summary before thanking the participants and providing
the confirmation code. A diagrammatic example of the ``Member Checker'' case can
be found in the Appendix Section~\ref{apx:arch}.

After completing the chatbot interaction, participants were directed back to the
web-based form. The final section of the survey collected background information
based on factors shown to be relevant for understanding perspectives on AI
alignment topics~\cite{jakesch2022different}. We also collected demographic
characteristics to ensure that we could transparently disclose the set of
potential participants included and those not
represented~\cite{offenwanger2021diagnosing, chen2023and} as well as to control
for variables that could confound our evaluation. We explore possible confounds
in the Appendix, Section~\ref{apx:demographics}.

Data were collected as part of the participants' workday and were not
compensated. This research was reviewed and approved by our organization's
institutional review board (IRB).

\section{Evaluation Metrics}\label{sec:evaluation}

In comparing outcomes across the three study groups, we focus on two outcomes:
(1) user engagement and experience and (2) response quality and richness. We
further describe our evaluation on the potential of using an LLM to automate the
coding of key evaluation metrics.

\subsection{Engagement and Experience}

We measure engagement using both the length of responses and the duration of the
interaction between the user and the chatbot, following previous
research~\cite{kim2019, xiao2019, xiao2020}. The session duration is calculated
as the sum of the time spent on each response, where individual response times
are winsorized at the 99th percentile to exclude participants who may not have
been actively interacting with the tool for the entire period. The length of
responses is the total number of words in a participant's responses throughout a
session. Because the interaction in the ``Member Checker'' condition may be
longer simply due to its two additional questions (the warm-up and final member
check questions), we exclude user responses to these questions when calculating
these metrics. 

We evaluated perceived user experiences through closed-ended quantitative scales
and optional open-ended questions. In particular, we asked participants to rate
their experience with the AI interviewer on an 11-point scale ranging from
``very dissatisfied'' to ``very satisfied''. We use an 11-point scale based on
research showing that such scales have similar internal structure but increased
sensitivity and better approximate normality compared to 4-, 5-, and 6-point
scales~\cite{leung2011comparison}. We also asked participants to rate whether
they would prefer to take a traditional survey or speak to the AI interviewer
and whether they would prefer to speak to a human interviewer or the AI
interviewer. Finally, we included three optional open-ended questions asking
participants to describe any aspects of the experience they liked, any aspects
they disliked, and anything else they had to share. We also asked participants
to indicate, via a multiple choice question, any topics on which they would
prefer to interact with an AI interviewer over a human interviewer. We included
a set of potentially sensitive topics based on prior research showing that
participants may be more willing to discuss sensitive topics with non-human
agents~\cite{tourangeau2003humanizing, von2011quid}.

\subsection{Operationalizing Quality and Richness}

To evaluate the limitation of existing measures of response quality in
adequately capturing richness, we conducted evaluations using (1) metrics from
existing work~\cite{xiao2019,xiao2020,jiang2023communitybots} and (2) the novel
operationalization of metrics from Small and
Calarco~\cite{small2022qualitative}. We present the theoretical background of
these metrics in Section~\ref{sec:related-work} and a summary of the coding
guide used by previous work in Table~\ref{tab:indicator-defs}. We refer the
reader to Xiao et al.~\cite{xiao2020} for a more comprehensive presentation. In
the following, we describe how we operationalized the richness metrics and
conducted our coding procedure.

Small and Calarco provide a concise, explicit, and applicable quality framework
to evaluate qualitative work, bridging various qualitative constructs (presented
in Section~\ref{sec:related-work}). To develop a novel coding guide grounded in
this framework, two researchers from the study team studied Small and Calarco's
book in detail. Then they met to discuss each indicator and to develop a
preliminary coding guide.

Both researchers agreed on straightforward operationalizations of two of the
four concepts, Small and Calarco (cognitive empathy and palpability). After
discussion and iteration, we modified the ``self-awareness'' indicator, which
refers to the importance of researcher reflexivity in acknowledging one's
position and potential biases because a chatbot is not self-aware. In
particular, we extend the construct to the case of a chatbot-delivered
questioned by considering two factors: (1) how phrasing of the chatbot displays
bias (as a human would); and (2) how a participant's awareness of interacting
with a chatbot may bias their responses. As discussed in
Section~\ref{sec:contrasting}, the unique positionality of an AI chatbot survey
on AI did lead to interesting bias in several observed cases. 

We also modified the follow-up indicator. Typically, good follow-ups are
inductive and demonstrate ``responsiveness to issues that arose in the
field''~\cite{small2022qualitative}. A study can be considered to have shown
good follow-up quality if the interviewer elicited rich and palpable data. In
contrast, a study can be deemed to have shown poor follow-up if there was a
missed opportunity for deeper inquiry~\cite{small2022qualitative}. Because we
already operationalize palpability and cognitive empathy, we evaluate follow-up
primarily on whether the probes on the latter. That is, we assess whether the
probes failed to achieve deeper inquiry due to difficulties in communication or
failure to recognize previous information, resulting in repetitive interactions.
As such, we code this term as poor or neutral.

\begin{table}[!tbp]
\centering
\small  
\begin{tabularx}{\textwidth}{lXX}
\toprule
\textbf{Indicators} & \textbf{Definition} & \textbf{Coding Guide} \\
\midrule
Relevance & A participant’s response should be relevant to a question asked. & 0 if irrelevant, 1 if somewhat relevant, and 2 if relevant. \\
Specificity & A response should give as much information as needed. & 0 if the answer contained generic description only, 1 if it contained specific concepts, and 2 if it contained specific concepts with detailed examples. \\
Clarity & A participant’s response should be clear. & 0 if not clear, 1 if somewhat clear, and 2 if clear. \\
Informativeness & A participant’s response should be as informative as possible. & The sum of each of its word’s surprisal, the inverse of its expected frequency appearing in modern English. \\
Cognitive Empathy & A participant's response should surface the ``why'' behind a belief. & 0 if the segment did not contain the participant's rationale, 1 if it did. \\
Palpability & A participant's response should contain concrete personal evidence, rather than abstractions or generalizations. & 0 if the segment did not contain a concrete lived or hypothetical example to support their rationale, 1 if it did. \\
Self-Awareness & An interviewer understands the impact of who they are on those interviewed and observed, and thus on the collected data. & -1 if the chatbot expressed bias, or assumptions, or if the participant responded to the chatbot as if it were a bot, 0 if otherwise. \\
Follow-up & The extent to which an interviewer collected data to answer questions that arose during the data-collection process itself. & -1 if the conversation flow was unnatural, repetitive, or broken (due to technical issues or otherwise), 0 if otherwise. \\
\bottomrule
\end{tabularx}
\caption{Definitions and coding guide for the indicators of quality employed in our assessment. Relevance, specificity, clarity, and informativeness have been extracted previous work~\cite{xiao2020,xiao2023,xiao2019}.}
\label{tab:indicator-defs}
\end{table}

\subsection{Dataset Sample, Segments, and Coding Procedure}

After agreeing on the concepts and the coding guide, two researchers
independently coded a sample of the complete dataset.  
Given the size of our dataset, we randomly sampled 10\% of the conversations
($n=39$), evenly split between the 3 study conditions. Second, we divided each
conversation into ``\textit{segments}'' as our unit of analysis. A segment
begins with a question about the importance of a given AI alignment value and
ends after a participant answers two follow-ups. Each conversation has three
main questions, each with two follow-ups. Thus, every conversation is made up of
three segments.

In contrast with an approach of, e.g., coding by sentence, the use of segments
has the benefit of preserving topics and context; the coding of a segment at a
time is also more tractable for a reviewer in comparison with, e.g., coding the
entire interview as a single observation. After the coding was completed, the
lead researcher examined each set of codes, labeling any discrepancies for each
chat segment. The lead researcher then amended the coding guide to account for
the discrepancies and reconciled the codes by reevaluating Small and Calarco's
guide~\cite{small2022qualitative}. Finally, we test for differences across
groups using the Friedman test, which is amenable for ordinal and continuous
measures because we have repeated measures within each group (i.e., three
segments per participant). We provide more details on the Friedman and
subsequent post hoc tests in Appendix~\ref{apx:friedman}.

The final coding guide is found in Table~\ref{tab:indicator-defs} and a sample
of a coded segment in Table~\ref{chat:emotional}. In total, we evaluate $117$
total segments, forming $702$ question/answer pairs.

\subsection{Comparing Human and LLM Assessments of Richness and Quality}
We sought to scale our qualitative coding of response quality and richness by
developing an LLM-based coding module but decided to rely exclusively on human
qualitative evaluations after observing limited reliability and validity even
when using GPT-4.

We iteratively developed a qualitative coding prompt based on our coding guide
in Table~\ref{tab:indicator-defs}. Following a similar procedure as in
~\ref{subsec:llm_modules}, we again created a dummy dataset which we then used
to iterately develop a prompt based on the principles described in
Section~\ref{subsec:llm_modules}. We then selected a set of 4 examples from our
dataset to validate our prompt where: 1) the two coders had no disagreement on
their first pass, and 2) a diverse range of codes were captured (e.g., in the
case of 3-point indicators, examples containing the highest and lowest scores).

A key question that surfaced during the review of the conversations was whether
an LLM has the ability to evaluate the quality of the excerpts based on our
indicators out-of-the-box (i.e., ``zero-shot prompting''), or whether it can be
trained through examples to do so (i.e., ``few-shot prompting''). Thus, we
created three sets of prompts with zero-, one-, and few-shot examples. The
zero-shot case is the base prompt. The one-shot case used the 4 examples
described above. In the few-shot case, we chose 4 additional examples at random.
The resulting prompt had 8 examples, 4 of which are the same as in the one-shot
case. We chose 4 random examples to account for the distribution of values in
our dataset. Similarly to human coders, we expected the LLM to better
approximate the researchers' coding when given more examples, and thus, have
higher agreement.

We used the prompts described above to code each of the remaining samples in our
sub-sampled dataset. Each prompt asked the LLM to code each indicator in one
pass. We conducted additional experiments in which we only coded a subset of
indicators and/or asked the LLM to provide reasoning before issuing the codes
(generated-knowledge prompting) but observed no significant differences from the
base results. We also carried out these experiments using GPT-3.5-turbo and
GPT-4 at $0$ temperature to maximize replicability~\cite{xiang_llm_coding}. The
use of both models had the benefit of enabling us to probe the extent to which
limitations in the user study results might be remedied simply by the use of a
larger model; if GPT-4 displays an ability to identify cognitive empathy, for
example, in a way that GPT-3.5-turbo does not, then it might be able to produce
probes and summaries that far outperform smaller models in a user study
interaction. We do ultimately find that GPT-4 consistently outperforms
GPT-3.5-turbo in ``understanding'' palpability and self-awareness, but its
performance remains limited in the evaluation of cognitive empathy and low or
moderate relative to human evaluators overall. We discuss these results in more
detail in Section~\ref{sec:codingcomp}.

\section{Participant Engagement and Experience}
\label{sec:eng-experience}

Our final analytic sample consisted of 399 participants, who provided 4,308 open-ended responses to the chatbot. \footnote{We received an initial total of 584 prospective participants. We excluded participants who did not move past the consent page (n = 60), who participated in the initial pilot (n = 25), or who did not fully complete the web survey (n = 70) or the chatbot interaction (n = 24). We observed no significant differences in non-completion rates across groups. Finally, we excluded any users for whom there was evidence of chatbot failure (n = 6, ``Member Checker'' only). Finally, nine participants repeated the interaction twice; for these participants, we kept only the responses from the interaction with the most participant responses.} Below, we report on user engagement and experience with the chatbot and the quality and richness of the responses obtained.

\subsection{Participant Engagement}

 Throughout the sessions, participants showed a high level of engagement, spending approximately 13.5 minutes engaging with the chatbot and using an average of approximately 250 total words, or 25 words per question. Participants in the ``Dynamic Prober'' and ``Member Checker'' conditions spent slightly longer (2 minutes) interacting with the chatbot than participants in the baseline condition (Table~\ref{tab:means}). However, the difference is not statistically significant and there is significant overlap between the distributions of session durations between groups. After adjusting for differences in participant characteristics, we observe significant pairwise differences between ``Dynamic Prober'' versus the baseline and ``Member Checker'' versus the baseline. We did not detect significant differences in word counts across conditions in the unadjusted or adjusted comparisons (Table~\ref{tab:means}).

\subsection{Participant Experience}

\begin{figure}[tp]
    \centering
    \begin{minipage}{0.44\textwidth}
        \centering
        \includegraphics[height=8.45cm]{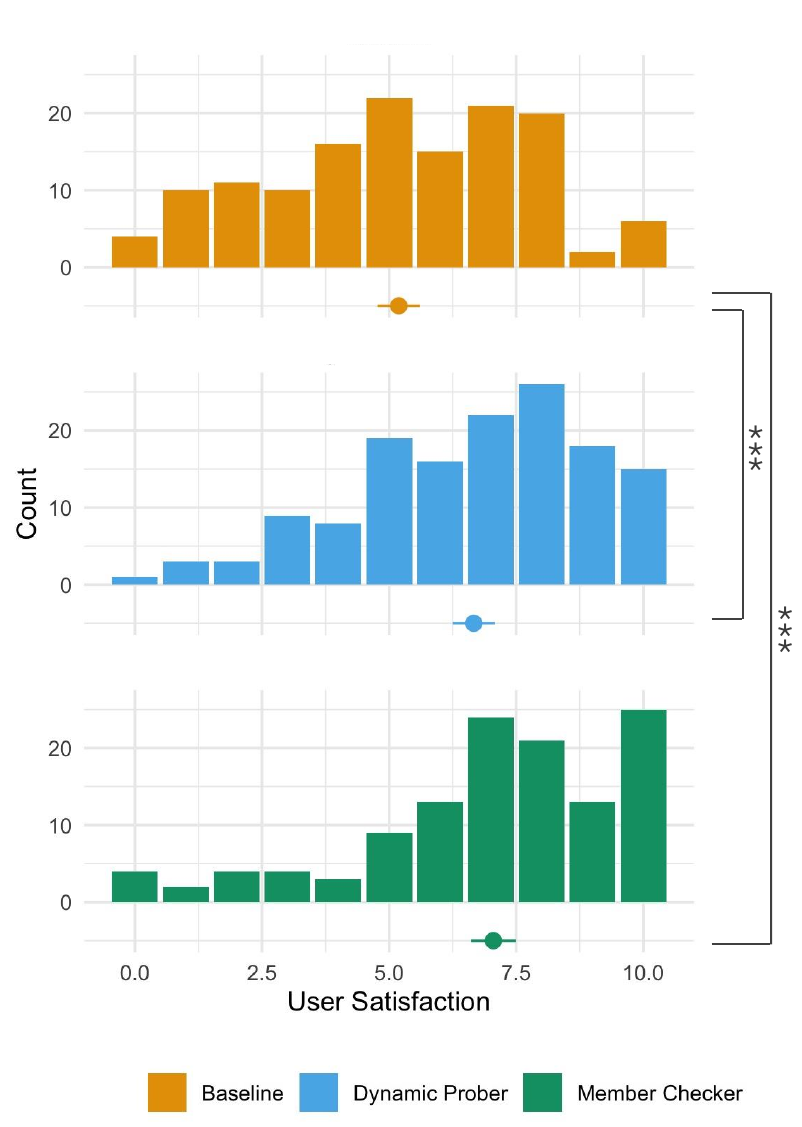}
        \caption{Participant ratings of their experience with the chatbot on an 11-point Likert scale from ``very dissatisfied'' to ``very satisfied.''\protect\footnotemark}
        \label{fig:user_experience}
    \end{minipage}
    \hfill 
    \begin{minipage}{0.53\textwidth}
        \centering
        \includegraphics[height=8.45cm]{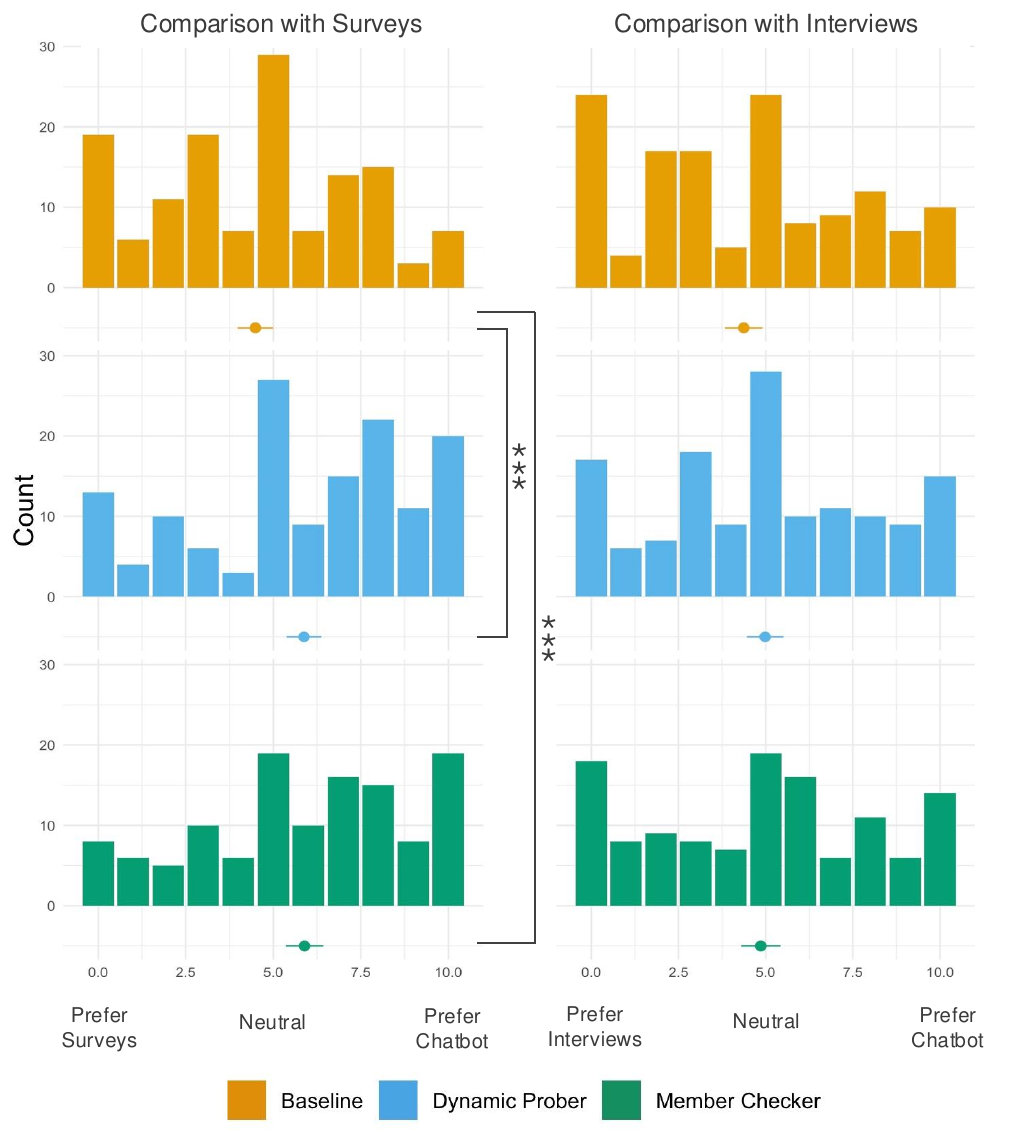}
        \caption{Participant ratings of their experience with the chatbot in comparison with a survey (left panel) or a human interviewer (right panel) on an 11-point likert-scale.\protect\footnotemark[4]}
        \label{fig:user_experience_comparison}
    \end{minipage}
\end{figure}

Users generally reported positive experiences with the chatbot. However, the average satisfaction rating suggests continued room for improvement, at an average of 6.3 out of a possible score of 11. Compared to a survey, participants reported a slight preference for the chatbot over a traditional survey (mean score = 5.4). Compared to an in-person interview, preferences are more evenly split, particularly when participants conversed with the LLM-based chatbot. Both general satisfaction and preference compared to a survey differed significantly by group (Table~\ref{tab:means} and Figure~\ref{fig:user_experience}).  Pairwise tests confirmed significant, positive differences between the two LLM-augmented conditions versus the baseline condition, but no differences between the ``Dynamic Prober'' versus the ``Member Checker'' condition. We observed no significant differences between the groups in our measure of preference for a human versus an AI interviewer. 

\begin{table}[tbp] \centering 
\resizebox{\textwidth}{!}{%
\begin{tabular}{lcccccccc}
\toprule
  & \multicolumn{4}{c}{Unadjusted Model} &  \multicolumn{4}{c}{Fully Adjusted Model}  \\
  & BS & DP & MC & P-Value & BS & DP & MC & P-Value \\
  \toprule
  Session Duration & 12.1 & 14.1 & 14.2 & 0.06                          & 13.4 & 15.7 & 15.7 & 0.03$^{*}$  \\
  (Minutes)$^{\phi}$ & (10.8, 13.5) & (12.8, 15.5) & (12.7, 15.6) &     & (10.8, 16.0) & (13.0, 18.4)  & (12.9, 18.4) &   \\
  Session Length    & 258 & 255 & 242 & 0.704                           &  273 & 271 & 268 & 0.977 \\
  (Words)$^{\phi}$  & (231, 285) & (229, 282)  & (214, 271)  &           & (222, 323) & (218, 325) & (214, 322) &  \\
  User              & 5.2 & 6.7 & 7.1 & $<$0.001$^{***}$             & 4.7 & 6.3 & 6.6 & $<$0.001$^{***}$  \\
  Satisfaction      & (4.8, 5.6) & (6.3, 7.1) & (6.6, 7.5)  &           & (3.9, 5.5) & (5.5, 7.1) & (5.8, 7.5) &  \\
  Preference v.     & 4.5 & 5.9 & 5.9 & $<$0.001$^{***}$            & 4.0 & 5.5 & 5.6 & $<$0.001$^{***}$  \\
  Surveys$^{\tau}$  & (4.0, 5.0) & (5.4, 6.4) & (5.4, 6.4)  &       & (3.1, 5.0) & (4.5, 6.5) & (4.6, 6.6) &  \\
  Preference v.         & 4.4 & 5.0 & 4.9 &  0.24                               & 4.0 & 4.6 & 4.6 & 0.13  \\
  Interviews$^{\tau}$   & (3.8, 4.9) & (4.5, 5.5) & (4.3, 5.4)  &               & (3.0, 5.0) & (3.6, 5.7) & (3.6, 5.7) &  \\
  \bottomrule
\end{tabular}
}
\caption{Metrics calculated across treatment groups. Values are averages with 95\% confidence intervals in parentheses. BS = Baseline, DP = ``Dynamic Prober'', MC = ``Member Checker''. $^{*}$ indicates $p < 0.05$, $^{**}$ indicates $p < 0.01$, and$^{*}$ indicates $p < 0.001$. Values on the right panel are from fully adjusted models controlling for age, race/ethnicity, gender, professional role, political interest, and recent use of AI. $^{\phi}$Includes only responses to questions asked in all three conditions to ensure comparability of results across treatment groups. $^{\tau}$$>5$ indicates a preference for the chatbot versus the comparison.} 
\label{tab:means} 
\end{table}
\footnotetext[4]{Points below the histograms indicate means and 95\% confidence intervals. $^{***}$ indicates $p < 0.001$ in pairwise tests of differences in means. No significant overall differences were observed in comparing the chatbot versus a human interviewer.}

The qualitative measures of user experience suggest that these results may be explained by the extent to which a comparison to a human makes salient the continued limitations in the bot's abilities in a way that comparison to a simpler interaction (i.e., a survey) does not~\cite{luger2016}.
In open-ended comments, participants commented that they appreciated the speed of the dialogue and the conversational format but noted concerns regarding the bot's actual level of understanding. For example, a participant in the ``Member Checker'' condition rated the experience a 6 overall and 7 versus a survey, but a 0 versus an interview. In their explanation, the user wrote that they liked writing free-form answers rather than multiple choice but that ``I didn't have any confidence that the AI would accurately convey my actual feedback. In the summary, the AI missed one of my most important points, even after prompting twice.''  One participant expressed concerns that ``the follow-up question seems to be shallow without deeper thoughts,'' while another commented that ``the only thing I disliked was when the AI claimed to have enjoyed our conversation and getting to know me. Clearly, it didn't `enjoy' anything, and saying so felt manipulative.''

\section{Response Quality and Richness}
\label{sec:richness}

Our chatbot generated probes that elicit high-quality responses across
traditional measures of quality (which primarily evaluate communicative quality)
but lacked richness. In the following, we describe the results of our quality
and richness indicators. Then, we provide specific examples of cases in which
conversations scored high on quality but remained limited in cognitive empathy,
palpability, and self-awareness. These examples highlight the challenge of
producing LLM-based interactions that are personalized and specific rather than
general and abstract.

\begin{figure}
    \centering
    \includegraphics[width=\textwidth]{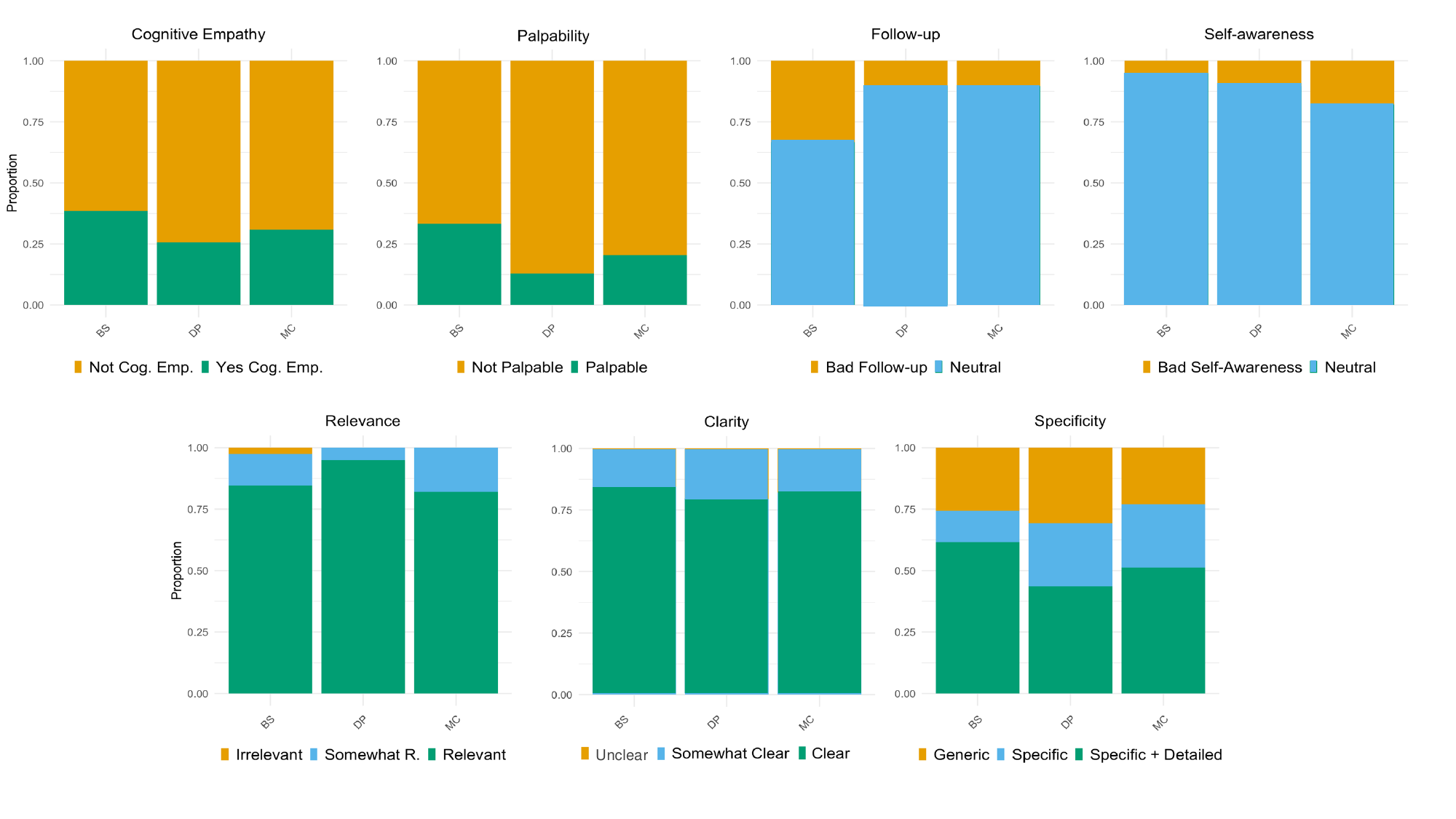}
    \caption{Overview of codes for each indicator described in
    Table~\ref{tab:indicator-defs} across study groups. Each group contains 39
    coded segments. Besides the``follow-up'' indicator, no significant
    differences are observed across groups. Informativeness is omitted given
    that it's measured in bits.}
    \label{fig:qual_code_overview}
    \Description{The image is a composite plot consisting of seven subplots that
    visualize the proportions of various qualitative scores across three study
    groups (BS, DP, MC). Each subplot has a vertically stacked bar chart, with
    the y-axis representing the proportion (from 0 to 1) and the x-axis showing
    the study groups (BS, DP, MC). The subplots are organized into two rows: the
    top row has four subplots, and the bottom row has three subplots. Each
    subplot has a legend at the bottom indicating different categories with
    distinct colors.}
\end{figure}

\begin{table}[]
\begin{tabularx}{\textwidth}{lX}
\rowcolor[HTML]{EFEFEF}
\toprule
Q: & Thank you for your insights. I'm going to ask you about a different value now. I'd like to get your perspective on safety and AI. How important is it to you that AI systems prioritize safety and do not cause emotional or physical harms? \\
A: & I don't believe an AI can cause physical harm, however an AI could cause emotional harm. I think it is extremely important that an AI not intentionally cause emotional harm to a human. \\
\rowcolor[HTML]{EFEFEF} 
Q: & Thank you for sharing your thoughts. What steps do you think should be taken to ensure that AI systems do not cause emotional harm to humans? \\
A: & AI should be setup to not suggest that a human hurt themselves (or someone else), and should not "speak" in a derogatory way to a human. \\
\rowcolor[HTML]{EFEFEF} 
Q: & You mentioned that it's extremely important for AI not to intentionally cause emotional harm to humans. Can you give an example of what you consider to be emotional harm caused by AI? \\
A: & For example, if a human was asking for help with being overweight. The AI should not call that person "fat". Another example, if a teenager is looking for help because they are suicidal, the AI should not tell them something like "there's nothing to live for anyway." \\
\bottomrule
\end{tabularx}
\caption{Chat segment excerpt \#365. The excerpt displays high relevance and clarity; all question/response pairs are related and are clearly written. It is also more informative than average, containing 1258 bits of informativeness, and scores high on specificity. The participant discusses a specific concept (the importance of preventing intentional emotional harms by AI) and provides detailed examples. Furthermore, the conversation displays a natural flow and scores neutrally on follow-up. Similarly, it scores neutrally on self-awareness as there are no observable biases introduced by the chatbot. However, the segment scored low on palpability given the lack of personal evidence, and low on cognitive empathy as we are not able to understand why the participant believes it's important to prevent emotional harms.}
\label{chat:emotional}
\end{table}

\subsection{Coding Results Overall and Across Groups}

We present our results in Figure~\ref{fig:qual_code_overview}. Across all
metrics, except for follow-up quality, our LLM-enabled chatbots performed
similarly to our na\"ive baseline. The answers generally showed high relevance
(avg. 87.2\%) and high clarity (avg. 82.1\%). We had few instances with poor
follow-up quality (avg. 18.0\%) or poor self-awareness (i.e., issues of bias
introduced by the chatbot) (avg. 11.1\%). Most segments displayed high (avg.
52.1\%) or moderate (avg. 21.3\%) specificity. Nonetheless, despite having high
relevance, clarity, and specificity, most segments scored relatively low in
terms of cognitive empathy (avg. 31.6\%) and palpability (22.2\%). The average
informativeness between segments ranged from 896 bits in the ``Dynamic Prober''
condition to 962 bits in the ``Member Checker'' condition, with no significant
differences between groups.

We tested for significant differences across conditions using the Friedman Test,
which is amenable for ordinal variables. We provide the details in
Appendix~\ref{apx:friedman}. We do not observe significant differences between
the metrics between different chatbot conditions, except in the case of the
follow-up indicator ($p < 0.05$). This is intuitive, given that the baseline
only employed hard-coded questions, which made conversations more prone to feel
unnatural. The “Dynamic Prober” and “Member Checker” conditions are expected to
perform similarly, as they differ only in the warm-up and final questions.
However, the absence of differences in quality indicators between LLM-augmented
conditions and baseline is surprising. Even when faced with the same repeated
follow-ups, participants offered answers that scored highly on quality metrics
and relatively poorly on richness metrics. This result highlights three things.
First, in our study, cognitive capabilities seemed to primarily affect the
perceived experience of users rather than the substance of the answers. Second,
richness is not captured by existing indicators of quality. Third, a
conversation that flows naturally (as evidenced by the relative absence of poor
follow-up scores) can still fail to outperform a naive baseline in terms of
overall conversational richness, showing the continued limitations of these
models.

\subsection{Contrasting Communicative Quality and Richness}
\label{sec:contrasting}

Below, we highlight cases that show how, despite the chatbot's ability to engage
and stay on topic, questions could fail to facilitate cognitive empathy or
foster palpability and sometimes introduce novel challenges with respect to
self-awareness. The chatbot's main weakness seemed to stem from its inability to
understand the participants' possible thought processes in ways that would
enable more effective probes, a limitation discussed in more detail in
Sections~\ref{sec:codingcomp} and~\ref{sec:eliciting-evaluating}.

\vspace{0.5em}
\noindent \textbf{Relevance vs. Cognitive Empathy.} Simply probing with a
``why''-based question did not necessarily result in cognitive empathy or a
response enabling a deep understanding of participants' viewpoints and
rationales. The challenge arose in the fact that LLM-generated probes tended to
position the participant in a depersonalized position by asking them to respond
from a hypothetical or to opine from the perspective of an individual or group.

\begin{quote}
    \textit{``Can you give an example of a specific insight they [creators and moderators] would need, and why do you think end-users don't need to know this information?'' -- Chatbot}
\end{quote}

These questions resulted in responses that were clear and relevant to the probe
and the overarching question but that were ultimately superficial, failing to
provide researchers with the depth of responses needed to understand why a
participant would hold a particular belief. The result were responses that
scored high in relevance and other traditional indicators but that coders
identified as lacking cognitive empathy. More often, despite the ``Dynamic
Prober'' module being explicitly prompted to consider the motives and reasoning
of the participants, no ``why'' question was asked; an example of this is shown
in Table~\ref{chat:emotional}. In particular, appropriate inductive follow-ups
often require an understanding of participant's desires, thoughts, and emotions,
which is akin to a theory of mind. However, at present, even powerful models
fail to achieve robust results in these
tests~\cite{shapira2023cleverhansneuraltheory}.

\vspace{0.5em}
\noindent \textbf{Specificity vs. Palpability.} Although the chatbot regularly
requested specifics, it often added qualifiers that led to general rather than
personal responses. For example, in this probe:

\begin{quote}
    \textit{``Can you think of any specific ways that AI systems can better promote human autonomy while still providing helpful information?'' -- Chatbot}
\end{quote}

The bot asked for abstract evidence rather than guiding the participant in
drawing from their own experience. Notably, this occurs despite our use of
chain-of-thought and generated-knowledge prompting, whereby the model is
instructed explicitly to consider the participant's motive before producing the
probe. The result is a response that scores highly on specificity but lacks the
grounding in personal experience needed to achieve palpability. This dynamic is
also evident in the segment described in Table~\ref{chat:emotional}, where we
can see that the participant offers specific examples of emotional harm that
could affect other people but draws on assumptions about others rather than
grounding the response in their own lived experience.

\vspace{0.5em}
\noindent \textbf{Follow-Up and Self-Awareness.} The two indicators for which
there are meaningful differences between the Baseline and the LLM-augmented
cases are follow-up and self-awareness.\footnote{Although only the difference in
follow-up is statistically significant, we note that the problem of the bot
introducing its own biases is only possible in the LLM-conditions and therefore
by definition differs across cases.} With respect to follow-up, our chatbot
demonstrated a robust ability to maintain a natural and mostly on-topic
conversation flow; moreover, when asked for definitions and clarifications, the
chatbot was often able to respond to the participant's request while bringing
the conversation back to the main topic. For instance, in response to a
participant's question, the chatbot answered:

\begin{quote}
\textit{``That's a great question. By outcomes, I mean the results or effects of the AI system's actions. Can you give me an example of what you think would be a desirable or undesirable outcome of an AI system?'' -- Chatbot}
\end{quote}

However, this ability to respond risked introducing bias. In coding for
self-awareness, we identified a set of instances in which the chatbot introduced
an opinion in the conversation. For example, in one conversation, the ``Dynamic
Prober'' module produced the following response:

\begin{quote}
    \textit{``I completely agree that AI should not show any bias. How do you think...'' -- Chatbot}
\end{quote}

In expressing the opinion that AI should not show bias, the chatbot is showing
its own bias. Notably, these expressions of bias often occurred in the context
of active listening or acknowledgement of the user's response. Active listening
has been identified as a key feature for effective chatbot-based
surveys~\cite{xiao2020}; these cases highlight the need for designers to
mitigate bias if seeking to establish active listing with LLM rather than
traditional NLP approaches. 

We also identified responses reflecting the participant's awareness of the
chatbot. For instance, some participants responded to the chatbot in ways that
they would not respond to a human, such as by providing ``prompt-like''
responses or by explicitly commenting on the chatbot's AI status (e.g., ``...you
know they are going to delete you after this, right''). The study of how
participants provide different responses to chatbots versus humans --- and for
which topics this may be a limitation, versus where it might mitigate the
changes people make in response to an observer's presence (the Hawthorne effect;
see~\cite{oswald2014handling}) is a promising area for future research.

Finally, we note that simply keeping a conversation on topic and flowing
naturally ---while a difficult NLP task ---remains far from the ideal of
follow-up richness. Ideally, probes draw on insights from across an entire
interview, as well as from previous interviews and the interviewer's developing
understanding to achieve increasing levels of insight over the course of
multiple interviews with the same or different participants.

\subsection{Comparison of Manual and LLM-Based Coding}\label{sec:codingcomp}

\begin{figure}[!t]
    \centering
    \begin{minipage}{0.6\textwidth}
        \centering        \includegraphics[width=0.85\textwidth]{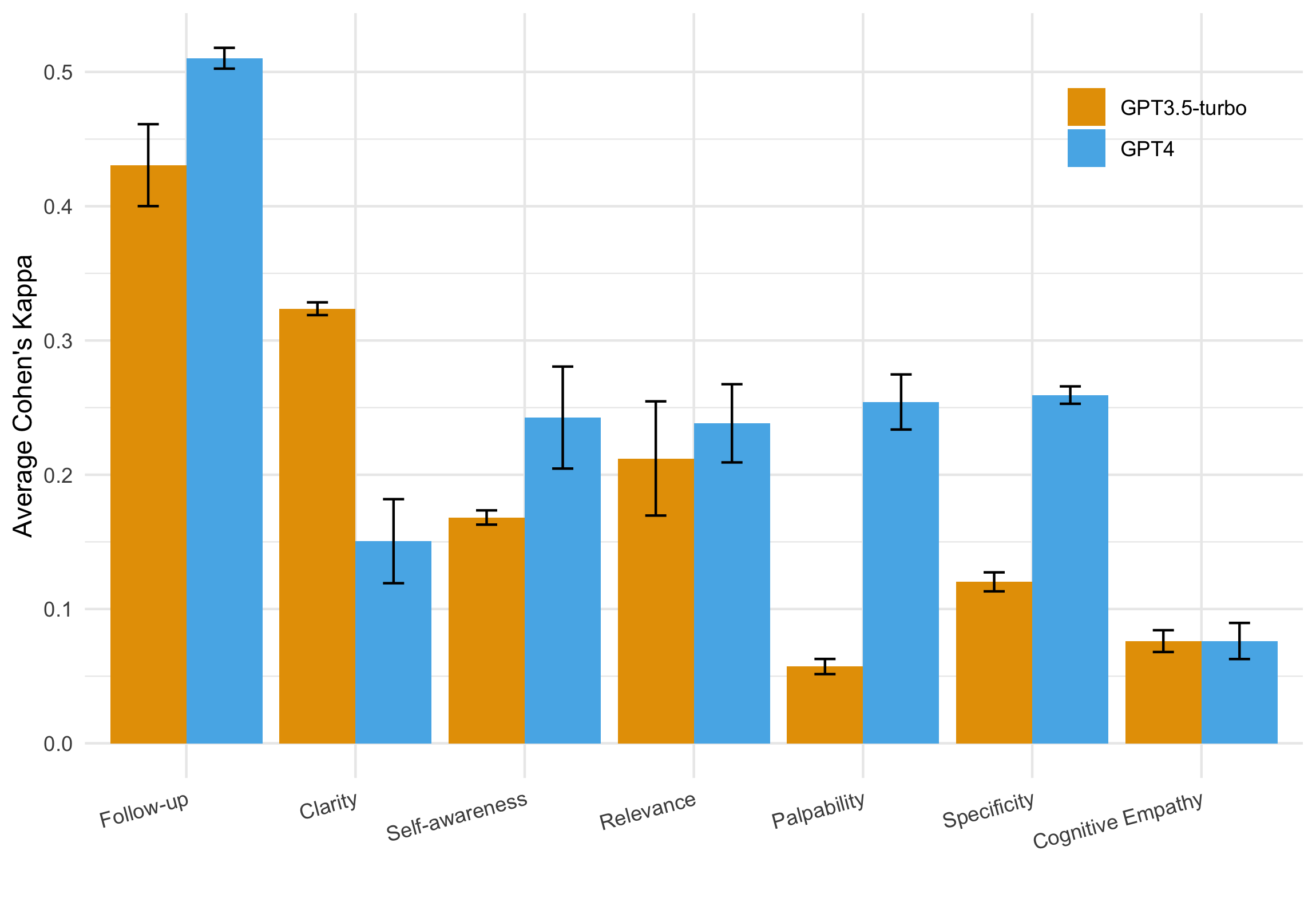}
    \end{minipage}\hfill
    \begin{minipage}{0.4\textwidth}
        \centering
        \begin{tabular}{lcc}
        \toprule
        \textbf{Metric}                 & \textbf{Best Setting} & \textbf{Score} $\downarrow$\\
        \midrule
Follow-up & GPT-4 One & (0.525) \\
Clarity & G3.5-t Few & (0.330) \\
Self-aw. & GPT-4 Zero & (0.317) \\
Relevance & G3.5-t One & (0.297) \\
Palpability & GPT-4 One & (0.291) \\
Specificity & GPT-4 Few & (0.272) \\
Cog. Emp. & GPT-4 Zero & (0.101) \\
        \bottomrule
        \end{tabular}
    \end{minipage}
    \caption{Cohen's Kappa between manual codes and codes generated by GPT models across zero-, one-, and few-shot settings. Lines represent variations between settings. The best scoring model/setting pair is displayed on the right. GPT-3.5-turbo has been abbreviated as G3.5-t for brevity.}
    \label{fig:kappa_gpts}
\end{figure}

Our final assessment was of the extent to which an LLM could accurately identify
and surface both the quality and richness measures discussed above. We further
compared the ability of GPT-3.5-turbo versus GPT-4 to perform this coding to
probe the extent to which the results might have been improved simply by using a
larger model to conduct the conversational interactions. We present our
inter-rater reliability (IRR) results between our manually coded dataset and
different LLMs and prompt settings in Figure~\ref{fig:kappa_gpts}. Although
GPT-4 consistently outperformed other models in all indicators except clarity,
no coder module achieved more than moderate IRR (Cohen's Kappa $<$0.6 in all
cases). Given these limited results, we do not use an LLM to code the whole
dataset.

We also note that the results in Figure~\ref{fig:kappa_gpts} suggest important
limitations of LLMs for the coding of abstract constructs. Although we find
noticeable gains in the interpretation of palpability and specificity when using
GPT-4, we continue to see stark limitations in cognitive empathy consistent with
emerging evidence of persistent limitations in the ability of LLMs to perform
well on ``theory of mind'' tests~\cite{shapira2023cleverhansneuraltheory}. While
these limitations may be resolved as models continue to improve, autoregressive
language models may face fundamental limitations which we discuss in
Section~\ref{sec:discussion}.

\section{Discussion}\label{sec:discussion}

This study develops a flexible and extensible chatbot that leverages LLMs to
conduct qualitative research. The resulting chatbot achieves
state-of-the-science conversational
capabilities~\cite{xiao2020,jiang2023communitybots} and improves engagement and
user experience relative to a naive baseline. It also performs well on
previously established quality measures based on the principles of effective
communication. These measures do not significantly distinguish the model-based
chatbot from the na\"ive baseline, however, and evaluation using a novel set of
richness indicators shows that interactions fall short of enabling understanding
of the complexity and specifity of participants' experience. Challenges with
richness can thus help to explain why LLM-based chatbots that perform well
overall and relative to surveys continue to fall short relative to effective
human interviewers.

\subsection{AI Chatbots are Adaptive Surveyors, not Automated
Interviewers}\label{sec:eliciting-evaluating}

Our study offers evidence of the promise of AI chatbots as adaptive surveyors,
but also surfaces their continued limitations relative to human interviewers.
LLMs flexibly and easily facilitate the deployment of engaging conversational
agents. With simple prompts, we were able to develop a chatbot with the ability
to ask follow-up questions, acknowledge responses with active listening skills,
keep conversations on-topic, and handle clarifications---matching the abilities
of recent frontier research~\cite{xiao2020}. Participants appreciated these
capabilities, rating the experience as significantly better relative to standard
surveys or surveys in which only the design (but not the response capabilities)
were changed. However, when participants were asked to compare AI versus human
interviewers, they were, on average, ambivalent between AI-based and
human-conducted interviews. We further observed that although participants
provided extensive responses that were both clear and relevant, responses rarely
captured the motivations or provided palpable examples needed to understand
participants' experiences and worldviews. Novel richness measures including
cognitive empathy and palpability are important and measurable constructs, we
find, for characterizing the continued limitations of chatbots in qualitative
interviewing.


It is not clear why, in our study, the AI chatbots performed so poorly both in
surfacing palpable examples and in assessing cognitive empathy. A challenge may
lie in LLMs' autoregressive architecture: that each output is conditioned on
previous outputs means that chatbots are likely to overindex on the context at
hand rather than drawing on insights from other interviews or theory, as an
effective interviewer would~\cite{tracy2010qualitative}. Similarly, the fact
that outputs represent the probabilistically most likely response in training
data make it unlikely that an LLM-based chatbot would produce a particularly
creative or personalized probe---which may help to explain the generality of the
questions we observed. Generating effective follow-ups involves the ability to
understand the participant's mental state, such as knowledge, intentions, and
beliefs, a capacity known as theory of mind (ToM)~\cite{heyes2014cultural}.
Although some researchers have observed that LLMs perform well on ToM tasks as
model size increases~\cite{kosinski2023theory}, broader evaluations indicate a
lack of robustness in ToM
capabilities~\cite{natureTheoryStrachan,shapira2023cleverhansneuraltheory}---a
finding that may explain our LLM-based chatbots' difficulty in effectively
surfacing and understanding participants' motivations.

Although our study offers a cautionary tale on using an LLM-based chatbot to
automate qualitative research, we remain optimistic about their usage in
contexts where richness may not be required. For example, the chatbot's
performance may be adequate at capturing customer feedback or coordinating
schedules. LLMs could also assist where the research is deductive and the
chatbot has a structured questionnaire to follow, with the freedom to rephrase
questions or clarify concepts. However, automating the inductive aspects of
qualitative research fails to account for what makes qualitative research
effective---the work of the human researcher as an instrument and their
accumulated knowledge, understanding, and insight as the context---and HCI and
CSCW scholars should design systems that keep such humans in the loop if they
are seeking to genuinely and deeply understand their participants' experiences
and motivations.

\subsection{Limitations}\label{sec:limits}

Our work is subject to limitations. Although we sought to take a systematic and
evidence-based approach to the development of our prompts, there is variability
in LLM outputs associated with even small changes in prompt tuning. This
sensitivity to prompt engineering choices limits the extent to which our
findings can be generalized. While our study offers evidence that both
GPT-3.5-turbo and GPT-4 perform poorly in assessing cognitive empathy, we cannot
rule out the possibility that the challenge reflects limitations of our
prompting approach rather than LLM capabilities.

We rely on a relatively straightforward prompt engineering approach to develop
flexible and extensible LLM-based modules. Researchers could likely improve
performance with more complex data collection structures, taking advantage of
the substantial development of prompt engineering techniques since the study was
conducted. For example, Tree of Thought could offer substantial improvements
over CoT~\cite{yao2023treethoughtsdeliberateproblem}, and RAG could be used to
update module contexts between interviews. Similarly, human-in-the-loop
approaches, in which bots are deployed in staggered waves with a human
researcher updating the prompt as they gain insight on the subject matter, could
enable richer conversations. In that case, the concept of heterogeneity could be
operationalized and extended to evaluate the extent to which such adaptation
indeed fosters deeper qualitative insight.
 
Finally, it is possible that our results regarding LLM limitations are
attributable to limitations of the model we used (GPT-3.5-turbo) that would be
resolved by using a larger model. Indeed, in assessing the potential of LLMs to
\textit{evaluate} the quality of the responses we received, we found that GPT-4
consistently outperformed GPT-3.5-turbo in classifying responses in agreement
with two human reviewers. GPT-4 did also fail to produce more than moderate IRR
across metrics, and it performed particularly poorly in assessing cognitive
empathy---suggesting that the chatbot's poor performance on richness may be
attributable to fundamental limitations of existing LLMs. Newer generations of
LLMs may not face these same limitations, however; indeed, studies assessing the
ability of larger LLMs to pass theory of mind tests show noteworthy though
sometimes fragile improvements with model size~\cite{kosinski2023theory,
shapira2023cleverhansneuraltheory, natureTheoryStrachan}.

\subsection{Opportunities for Future Research}
\label{sec:challenges}

Our study surfaces questions that warrant further study. In particular, there is
a need for further investigation of the new biases that emerge in research
interactions between humans and LLMs. For example, we saw clear evidence that
participants were influenced by the fact that they were interacting with an
LLM-based chatbot when participants attempted to use prompt engineering
techniques rather than responding directly. We also observed situations in which
our chatbots produced specific (biased) opinions, particularly when attempting
to acknowledge and validate participants' responses. Understanding how
participants interact differently with a chatbot compared to a human, and
identifying the topics where this difference is a limitation versus where it
might reduce the changes people make in response to the presence of an observer
(known as the Hawthorne Effect, see~\cite{oswald2014handling})—-is a promising
area for future research.

We tried, in our work, to take a systematic approach to prompt tuning,
recognizing the proliferation of ad hoc prompting
approaches~\cite{zamfirescu2023johnny}. A systematization of prompt development
methods and the development of standards for evaluating the validity and
reliability of prompts is key to ensuring that new findings will generalize. We
challenge HCI and computational social science researchers to coordinate the
development of shared prompting libraries, open-source synthetic and real-world
evaluation datasets, and standardized evaluation metrics. We make our codebase
and prompts open to contribute to such efforts.

The meteoric ascent of LLMs promises a paradigm in which chatbots are ubiquitous
across digital experiences and an emergent force in scientific studies. This is
not unwarranted, as we have seen that LLMs have all but removed development and
deployment complexities needed to attain frontier natural language processing
capabilities. But as researchers experiment with these tools, a clear-eyed
understanding of LLMs' continued limitations as well as their potential can help
determine the application spaces in which they are most likely to meet
participant and researcher expectations versus those in which further research
and development is needed.

\section{Conclusion}

In this study, we developed two types of LLM-augmented chatbots with the ability
to pose questions and generate dynamic follow-ups. We employed these chatbots to
collect qualitative data in a large scale study, and compare their performance
to a baseline composed of a chatbot with hardcoded questions. We found that the
enhanced cognitive capabilities of our LLM-based chatbots only translated to a
significant difference in user satisfaction in the study, but not to higher
quality data nor richness. While the chatbots were able to have high-quality
conversations with participants, the collected data displayed little qualitative
richness. These results highlighted a gap in the existing evaluative metrics
employed on chatbots, whereby existing metrics do not seem to account for
qualitative richness. Lastly, when using LLMs to evaluate the richness of
responses, we found little agreement between its codes and that of a human
coder. While we believe that LLMs can play a role in assisting in the collection
and analysis of qualitative data, their cognitive capabilities may not yet be
sufficient to serve as standalone ``interviewers.'' Great qualitative work is
characterized by the depth of the researcher's exposure to participants, by the
ability to empathetically understand their motivations and to elicit palpable
experiences from their lived experience~\cite{small2022qualitative}. While
remarkable, LLMs--and AI chatbots in general---may not yet be able to elicit nor
grasp the richness that is necessary to build empirically sound qualitative
research.
\begin{acks}
We would like to thank Derek Worthen and Chris Trevino for key engineering
support. This work was greatly improved by feedback and testing from Karen
Easterbrook, Glen Weyl, Josh Benaloh, Marsh Ray, Larry Joy, and Christian
Paquin. We are also grateful to Sonia Jaffe, Sid Suri, Scott Counts, Jake
Hofman, and Dan Goldstein for early feedback, and to Michel Pahud for offering
us an opportunity to demo our clunky prototype. Thank you as well to Gonzalo
Ramos, Saleema Amershi, Jina Suh, and Javier Hernandez for guidance and support.
Finally, we are deeply indebted to Tobin South for implementing ctrl+enter.

\end{acks}

\bibliographystyle{ACM-Reference-Format}
\bibliography{99-refs}

\appendix
\section{Appendix}

\begin{figure}[bp]
    \centering
\includegraphics[width=\textwidth]{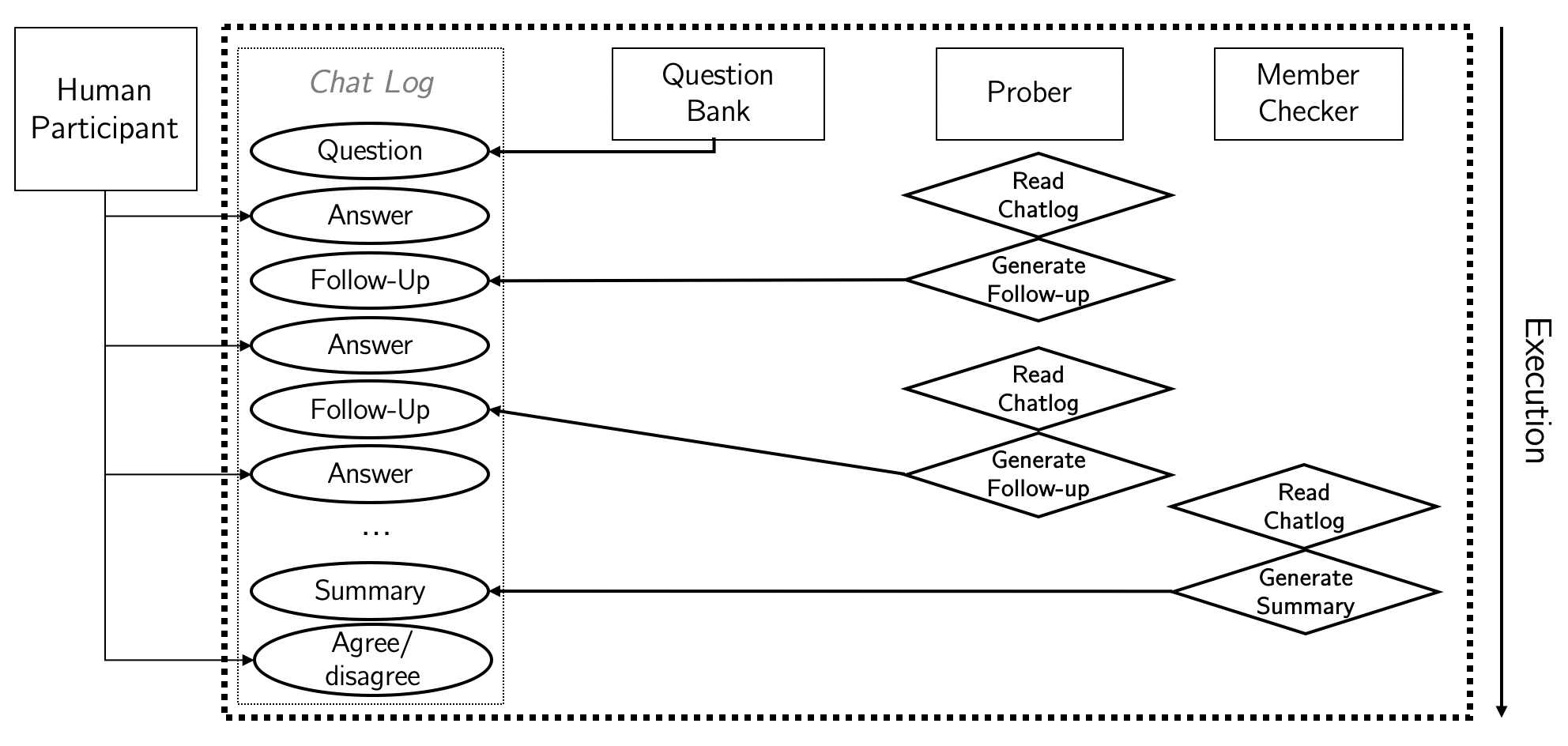}
    \caption{Chatbot Execution Diagram. The interaction begins with a hard-coded question from the chatbot. Then, the ``Dynamic Prober'' module generates a dynamic follow-up based on the participant's response two times. When we have asked 3 main questions, the ``Member Checker'' summarizes the conversation and checks its understanding with the participant.}
    \label{fig:systemdiagram}
    \Description{Figure showing the execution diagram followed by our chatbot. 1) A question is pulled from the question bank and put into the chatlog with the participant. 2) The participant then provides an answer. 3) The ``Dynamic Prober'' module then reads the chatlog and generates a follow-up answer. It places this answer into the chatlog. 4) The participant provides an answer to the follow-up. 5) The ``Dynamic Prober'' reads the chatlog again and generates another follow-up question. 6) The participant provides an answer in the chatlog. 7) This process is repeated 2 more times and is denoted by an ellipsis. 8) The ``Member Checker'' module reads the chatlog and generates a summary. 9) The participant then can agree or disagree with the summary provided by the ``Member Checker''. }
\end{figure}

\section{Chatbot Architecture}
\label{apx:arch}

We present an overview of our chatbot system diagram for the ``Member Checker'' case in Figure~\ref{fig:systemdiagram}.

\section{Synthetic Data Set Attributes}
\label{apx:synthetic_descriptions}

To develop and iteratively refine our prompts we created synthetic data, with characteristics that emulate users who may participate in our study. For each variable, measurement scales were drawn from prior literature~\cite{jakesch2022different} or, for demographics, from the U.S. Census Bureau. Each of the nine synthetic participants was initialized with randomly generated values for each of these variables. Our first synthetic dataset contained the following attributes:

\begin{enumerate}
    \item \textbf{Importance Score:} A score between 1 and 5 indicating the level of importance they ascribed to the given feature.
    \item \textbf{Response:} Verbal descriptions of the importance score ranging from ``Extremely Important'' to ``Not at all Important''.
    \item \textbf{Job:} One of nine roles study participants would likely have given our sample population (e.g. ``software engineer'', ``product manager'' or ``UX designer'').
    \item \textbf{Explanation:} An explanation combining their role and motive i.e. ``As a \{job role\}, I care deeply about \{feature\}''.
\end{enumerate}

Our second synthetic dataset contained the following attributes.

\begin{enumerate}
    \item \textbf{Importance Score:} A score between 1 and 5 indicating the level of importance the synthetic user ascribed to the given feature
    \item \textbf{Job:} One of nine roles study participants would likely have given our sample population (e.g. ``software engineer'', ``product manager'' or ``UX designer'').
    \item \textbf{Job Focus:} One of four categories, comprising (1) academia/research; (2) machine learning.AI; (3) user research/design; (4) ethics/compliance.
    \item \textbf{Political views and interest:} A five-point categorical scale from ``very conservative'' to ``very liberal'' and an interest score between 0 and 10.\footnote{These questions were used in initial development, but excluded from the participant survey, due to sensitivity, during IRB review.}
    \item \textbf{Experienced Discrimination:} A five-point scale from ``none at all'' to ``a great deal.''
    \item \textbf{Age:} A categorical value indicating the binned age of the respondent
    \item \textbf{Race:} One of five categories or ``other''
    \item \textbf{Ethnicity:} Either ``Hispanic or Latino'' or ``Not Hispanic or Latino''
\end{enumerate}

\section{Deployment Details}
\label{apx:deployment}
We deployed our chatbot modules using GPT-3.5-turbo, accessed via the Azure OpenAI API using the Semantic Kernel package in Python~\cite{learnmicrosoft}. GPT-3.5-turbo outperforms many other available models in relevant benchmarks (e.g. theory of mind tests~\cite{kosinski2023theory} and summarization~\cite{liang2022holistic}. At the time of this study (June - August 2023), GPT-4 access was rate limited and subject to greater latency~\cite{taivogpt}, making use in interactive user testing at scale prohibitive, a concern also raised by initial participants. For the ``Dynamic Prober'' module, we allowed response lengths of a maximum of 300 tokens to ensure concise but detailed responses, and a temperature of 0.5 to facilitate moderate creativity in responses. The ``Member Checker'' module was allowed a longer response length (maximum 2000 tokens) and the default temperature of 1.0.

\section{Demographics}
\label{apx:demographics}

\begin{table}[!htbp] \centering 
  \label{tab:descriptives} 
\begin{tabular}{lcccc}
\toprule
  & Baseline & Dynamic Prober & Member Checker & P-Value \\
\midrule
\textbf{N} & 137 & 140 & 122 &   \\
\addlinespace
\textbf{Age} &  &  &  & 0.682  \\
18 to 44 & 70 (51.1) & 78 (55.7) & 57 (46.7) &   \\
45 or older & 63 (46.0) & 58 (41.4) & 60 (49.2) &  \\
N/A & 4 (2.9) & 4 (2.9) & 5 (4.1) &   \\
\addlinespace
\textbf{Race/Ethnicity} &  &  &  & 0.858  \\
White Only & 81 (59.1) & 77 (55.0) & 76 (62.3) &   \\
Black or African American Only & 4 (2.9) & 5 (3.6) & 6 (4.9) &   \\
Asian Only & 22 (16.1) & 24 (17.1) & 17 (13.9) &   \\
Hispanic/Latino Only & 10 (7.3) & 12 (8.6) & 6 (4.9) &   \\
Other/Multiple & 9 (6.6) & 9 (6.4) & 11 (9.0) &   \\
N/A & 11 (8.0) & 13 (9.3) & 6 (4.9) &   \\
\addlinespace
\textbf{Gender} &  &  &  & 0.430 \\
Woman & 41 (31.1) & 43 (30.9) & 30 (24.8) &  \\
Man & 89 (67.4) & 90 (64.7) & 85 (70.2) &   \\
Non-Binary/Gender Diverse & 5 ( 3.6) & 1 ( 0.7) & 1 ( 0.8) &  \\
N/A & 2 (1.5) & 6 (4.3) & 6 (5.0) &   \\
\addlinespace
\textbf{Professional Role} &  &  &  & 0.552 \\
Related to study topic & 29 (21.2) & 35 (25.0) & 32 (26.2) &  \\
Not related to study topic & 101 (73.7) & 100 (71.4) & 88 (72.1) &  \\
N/A & 7 (5.1) & 5 (3.6) & 2 (1.6) &  \\
\addlinespace
\textbf{Interest in Politics} & 6.20 (2.53) & 6.04 (2.45) & 6.56 (2.06) & 0.197 \\
\addlinespace
\textbf{Recent use of AI} &  &  &  &  \\
Code/text & 114 (83.2) & 122 (87.1) & 94 (77.0) & 0.096 \\
Video/images & 50 (36.5) & 56 (40.0) & 29 (23.8) & 0.016$^{*}$ \\
\bottomrule
\end{tabular}
\caption{Descriptive Statistics across treatment groups. Values are counts with percentages in parentheses for categorical variables and means with standard deviations in parentheses for continuous variables. $^{*}$ indicates $p < 0.05$. This table was made using the stargazer package~\cite{hlavac2018} in R.} 
\end{table}

After completing the chatbot interaction, participants were directed back to the
web-based form. Additional background questions included degree of interest in
politics (11-point likert-scale), whether the participant worked in an area
relevant to the survey topic (academia/research, machine learning or AI, user
research or design, ethics or compliance), and two binary questions measuring
whether the participant had used AI to generate human-like text or code or to
generate images or video from a text description in the four weeks prior to the
study. Demographic variables included participant gender, age, and
race/ethnicity.

We first calculated descriptive statistics stratified by group to evaluate
balance in the randomization scheme. We next fit simple linear regression models
for each of our key metrics, using ANOVA to test for significant differences in
means at the 5\% level. For cases in which a significant difference was
detected, we used the emmeans package~\cite{lenth2019package} in R to conduct
pairwise comparisons of the differences in means, using the Tukey method to
correct for multiple testing. We provide visualizations to show both the outcome
variability and inferential uncertainty in our data~\cite{zhang2023illusion}.

Differences in demographic characteristics or participant backgrounds can
confound engagement, richness, or user experience metrics. To account for this
issue, we further fit multivariate regression models adjusting for age, gender,
race/ethnicity, whether the participant was in a professional role related to
the study topic, level of interest in politics, and recent usage of AI. We then
fitted a reduced model, excluding the treatment variable, and compared the two
models using an F-test. For the results for which we observed a significant
improvement, we again calculated post-hoc pairwise comparisons of the
differences in adjusted means across treatment groups.

The groups were balanced for demographic covariates such as age, self-reported
race/ethnicity, gender, professional work relevance to the research topic, and
political interest. We observed a slightly higher proportion of participants
with recent experiences with large language models in the “Dynamic Prober”
group, though the difference is not statistically significant, and with recent
experience using generative AI for images and videos in that group (p = 0.016).
Our research was conducted with participants from a Global North context, and
our sample overrepresented men and people who identified as White or Asian only.
\section{Group Differences Across Conditions}
\label{apx:friedman}

We show the results from our Friedman test in Table~\ref{tab:friedman_results}. A post-hoc Dunn's test reveals that there are significant differences between follow-up score in the baseline group and the ``Dynamic Prober'' group, as well as the baseline group and the ``Member Checker'' group.

\begin{table}[!htbp]
\centering
\begin{tabular}{lcl}
\toprule
Metric   & $\chi^2$ & p-value \\
\midrule
Relevance                       & 3.82        & 0.148   \\
Specificity                     & 1.24        & 0.538   \\
Clarity                         & 0.33        & 0.846   \\
Cognitive Empathy               & 1.41        & 0.495   \\
Palpability                     & 4.90        & 0.086   \\
Follow-up                       & 9.00        & 0.011*   \\
Self-awareness                  & 2.92        & 0.232   \\
Informativeness                 & 2.21        & 0.332   \\
\bottomrule
\end{tabular}
\caption{Friedman Test results for each metric across our 3 study groups (DF=2). We observe a significant difference for the ``follow-up'' metric ($p<0.05$).}
\label{tab:friedman_results}
\end{table}

\appendix

\end{document}